\documentclass[preprint,10pt]{elsarticle}

\usepackage{geometry}
\usepackage{graphicx}
\usepackage{amsfonts}
\usepackage{amsmath}
\usepackage{amssymb}
\usepackage{amsthm}
\usepackage{bm}
\usepackage{subfig}
\usepackage{float}
\usepackage{xcolor}
\usepackage{lineno}
\usepackage{natbib}

\usepackage[
colorlinks=true,
linkcolor=blue,
citecolor=blue,
urlcolor=blue
]{hyperref}

\newcommand{\sech}{\operatorname{sech}}
\newcommand{\sn}{\operatorname{sn}}
\newcommand{\cn}{\operatorname{cn}}
\newcommand{\dn}{\operatorname{dn}}
\newcommand{\ii}{\mathrm{i}}

\begin{document}
	
	\begin{frontmatter}
		
		\title{
			Phase Space Reorganization and Traveling Wave Emergence
			Driven by Non-Kerr Effects in Nonparaxial Optical Media
		}
		
		\author[1]{Naresh Saha}
		\ead{saha.naresh92@gmail.com}
		
		\author[2,3]{Nirmoy Kumar Das}
		\ead{nirmoymath@gmail.com }
		
		\author[3]{Ashoke Das}
		\ead{ashoke.avik@gmail.com }
		
		\author[4,5]{Arnob Ray\corref{cor1}}
		\ead{arnobray93@gmail.com}
		
		\address[1]{
			Department of Mathematics, School of Engineering,
			Dayananda Sagar University,
			Bengaluru, Karnataka 562112, India
		}
		
		\address[2]{
			Department of Mathematics, Kaliyaganj College,
			Uttar Dinajpur, West Bengal 733129, India
		}
		
		\address[3]{
			Department of Mathematics, Raiganj University,
			Uttar Dinajpur, West Bengal 733134, India
		}
		
		\address[4]{
			Department of Civil Engineering,
			Indian Institute of Technology Gandhinagar,
			Gandhinagar, Gujarat 382055, India
		}
		\address[5]{
			Department of Mathematics,
			SRM Institute of Science \& Technology,
			Kattankulathur, Tamil Nadu 603203, India
		}
		
		\cortext[cor1]{Corresponding author}
		
		\begin{abstract}
			In this article, the nonlinear Helmholtz equation with non-Kerr
			nonlinearity, such as self steepening and self frequency shift, is considered.
			A traveling wave transformation is applied, and the extended nonlinear
			Helmholtz equation is reduced to a Hamiltonian dynamical system. Then, the
			reduced Hamiltonian system is analyzed by classification of equilibrium
			points, phase space analysis, and the construction of exact wave solutions.
			The relationship between the reduced dynamical coefficients and the original
			physical parameters is further established through a parameter space
			analysis. It is shown that self steepening directly modifies the reduced
			dynamics, whereas self frequency shift acts through the compatibility
			condition for the real traveling wave reduction. Together, these non-Kerr
			effects reshape the phase space geometry and traveling wave structure.
			Localized and periodic traveling waves are obtained, with their existence
			determined by the balance among dispersion, nonparaxiality, Kerr nonlinearity,
			and non-Kerr effects. Furthermore, a periodically forced version of the
			reduced system is examined to study the transition from regular to irregular
			dynamics. It has been observed that external forcing can induce complex
			oscillatory behavior. Bifurcation analysis, time series evolution, phase
			space analysis, largest Lyapunov exponent, and Poincar\'e section demonstrate
			the emergence of quasiperiodic and chaotic responses under sufficiently
			strong forcing. All analytical branches are verified through full-equation
			residual evaluation, while a few selected branches are additionally examined
			through direct numerical propagation and robustness tests under complex
			Gaussian perturbations. The results show that self steepening directly
			renormalizes the effective nonlinear dynamics, whereas self frequency shift
			restricts the admissible real-envelope traveling wave manifold.
		\end{abstract}
		
	\end{frontmatter}
	
	
	\begin{quotation}
		This study establishes a direct connection between non-Kerr optical
		effects and the geometry of the reduced Hamiltonian system. Self steepening
		(SS) modifies the effective nonlinear coefficient, whereas self frequency
		shift (SFS) constrains the compatibility manifold required for real envelope
		traveling waves. The resulting phase space organization determines the
		existence of localized and periodic solutions and governs their response
		under external forcing.
		Here, we consider the nonlinear Helmholtz model incorporating these
		non-Kerr nonlinearities, SS and SFS. By applying a traveling wave reduction
		and a dynamical systems analysis, we show that these non-Kerr nonlinear
		effects do more than just slightly modify the system. They change the
		structure of the phase space, affect the equilibrium points, and modify
		the bifurcation behavior of the system. Because of this, the model can
		generate different types of traveling wave solutions, such as localized
		waves, periodic waves, as well as more complex oscillations under external
		forcing. The central outcome of this work is to show that SS and SFS play
		an important role in shaping the system dynamics of our proposed model,
		rather than acting as small corrections to previously known solutions.
	\end{quotation}
	
	\section{Introduction}
	\par Nonlinear wave propagation in dispersive media is a foundational subject in modern mathematical physics and nonlinear optics, since it furnishes a canonical framework where localization, modulation, bifurcation and rich spatiotemporal complexity arise from the subtle interaction between dispersive spreading and nonlinear self interaction \cite{Ablowitz2011,Sulem1999,KivsharAgrawal2003}.
	Among the many model equations used in this context, the nonlinear Schr\"odinger equation (NLSE) occupies a foundational position, since it describes the evolution of slowly varying optical envelopes in weakly nonlinear and weakly diffracting or dispersing media \cite{Agrawal2019,HasegawaKodama1995,Sulem1999}. The NLSE derives from Maxwell's equations under the paraxial (or equivalently slowly varying envelope) approximation, which requires that the optical beam is at once much broader than its carrier wavelength, of sufficiently low intensity, and propagating predominantly along the reference axis \cite{Lax1975,Agrawal2019,Barton1989}. When these assumptions are violated, the optical field leaves the paraxial and slowly varying envelope regime, with beams exhibiting spatial nonparaxiality and pulses additionally reflecting the breakdown of the slowly varying temporal envelope approximation \cite{chamorro1998non,Christian2009}. Interest in such regimes goes back to the pioneering work of Lax et al. \cite{Lax1975}, who examined the transition from Maxwell's equations to paraxial wave optics and thereby clarified the limits of the paraxial approximation.
	
	\par A natural framework for this broader regime is provided by the nonlinear Helmholtz (NLH) model \cite{ChamorroPosada2002,laine2000self}. In contrast to paraxial envelope equations, the NLH formulation retains the longitudinal correction that becomes important when beam geometry or propagation direction is no longer asymptotically close to the paraxial limit \cite{ChamorroPosada2002,laine2000self,SanchezCurto2007,saha2022dipole}. This feature has motivated substantial work on scalar and coupled Helmholtz systems, including the derivation of exact analytical soliton solutions for focusing and defocusing Kerr media, as well as extensions involving power-law and polynomial nonlinear responses \cite{ChamorroPosada2002,Christian2007,christian2010bistable,christian2012helmholtz,christian2007helmholtz,christian2009bistable}. The effect of nonparaxiality on the amplitude, pulse width, and propagation speed of solitary waves has also been investigated \cite{tamilselvan2016nonparaxial}. The modulational instability of the cubic-quintic NLH system is analyzed in \cite{Tamilselvan2019}, where chirped elliptic and hyperbolic solitary wave solutions are reported. Furthermore, the cubic coupled NLH equation incorporating self steepening (SS) and self frequency shift (SFS) effects is examined in \cite{Saha2021,saha2022chirped}.
	
	\par At the same time, realistic ultrashort pulse and strongly nonlinear optical propagation is often influenced by non-Kerr nonlinearity effects \cite{Agrawal2019,Francois1991}. Two such effects are particularly important, i.e., SS and SFS\cite{deOliveira1992, Gordon1986,Mitschke1986, Trippenbach1998}. SS introduces intensity dependent temporal distortion and can break the symmetry of the envelope profile, while SFS alters the phase and spectral structure of the pulse through higher order nonlinear coupling \cite{deOliveira1992,Gordon1986,Trippenbach1998}. In many studies, these effects are incorporated perturbatively into NLSE models, where they are often viewed as corrections to an already established pulse family \cite{Agrawal2019,Francois1991,Trippenbach1998}. However, when these mechanisms are incorporated within a nonparaxial Helmholtz framework, their influence becomes substantially more profound, as they can renormalize the effective reduced coefficients, transform the equilibrium structure of the traveling wave dynamics, deform invariant manifolds in phase space, and thereby reorganize the very families of admissible waves \cite{Ablowitz2011,Saha2021,Tamilselvan2019,Strogatz2018}.

	\par This observation reveals the main research gap addressed in the present work. Although previous literature has separately established the importance of nonparaxial Helmholtz propagation, exact Helmholtz solitons, and non-Kerr nonlinearity such as SS and SFS, a unified dynamical systems analysis showing how these mechanisms simultaneously reorganize the phase space geometry and traveling wave structure of an extended scalar nonlinear Helmholtz model is still lacking \cite{ChamorroPosada2002, Christian2007, Tamilselvan2019, Saha2021}. In particular, it remains unresolved whether, when SS and SFS are incorporated alongside the Helmholtz longitudinal correction, they simply perturb previously known wave states or instead fundamentally reconstruct the reduced Hamiltonian landscape, bifurcation structure, and the very class of admissible localized, periodic, and singular solutions \cite{ur2024bifurcations,iqbal2025exploring,rahaman2025bifurcation,hamad2025bifurcation,alraqad2024investigating}.
	
	\par Motivated by the above research gap, we investigate an extended NLH equation describing nonparaxial optical pulse propagation in the presence of Kerr and non-Kerr nonlinear effects such as SS, SFS, together with the longitudinal nonparaxial correction \cite{ChamorroPosada2002,Tamilselvan2019,Saha2021}. By employing a symmetry guided traveling wave ansatz, the governing NLH equation is reduced to a planar Hamiltonian dynamical system, which is subsequently analyzed through equilibrium classification, phase space geometry, bifurcation analysis, explicit quadrature, and periodically forced dynamics \cite{Ablowitz2011,Strogatz2018}. This reduced formulation provides a geometrically transparent and analytically tractable framework that compresses the combined effects of dispersion, carrier modulation, Kerr nonlinearity, SS, and nonparaxiality into a small set of effective dynamical parameters. Consequently, the existence and nature of traveling wave states can be understood directly from equilibrium configurations, invariant level sets, separatrices, and families of bounded orbits \cite{Ablowitz2011,Strogatz2018}. In particular, the reduced formulation establishes a direct connection between the original physical coefficients and the effective dynamical parameters, while quantitative bifurcation and chaos diagnostics together with numerical stability analysis provide further validation of the resulting traveling wave solutions. The analysis further shows that SS directly modifies the effective cubic coefficient of the reduced Hamiltonian system, whereas the SFS contribution acts through the compatibility condition that determines the admissible real traveling wave reduction. Together, these non-Kerr effects govern the organization of the reduced phase portrait and the emergence of localized solitary waves, periodic wave trains, transition-type traveling wave profiles, and increasingly complex oscillatory dynamics under external forcing. The principal contribution of this work is therefore not merely the incorporation of higher order nonlinear effects into a known optical model, but the demonstration that these effects fundamentally reorganize the underlying Hamiltonian structure of the nonparaxial system, thereby providing a systematic framework for relating higher order physical mechanisms to qualitative changes in phase space geometry, bifurcation structure, coherent wave formation, and nonlinear propagation dynamics \cite{GuckenheimerHolmes1983,ur2024bifurcations,iqbal2025exploring,rahaman2025bifurcation,hamad2025bifurcation,alraqad2024investigating}.
	
	\par The paper is organized as follows. In Sec.\ \ref{sec2}, we introduce the extended nonlinear Helmholtz model and derive the traveling wave reduction together with the associated compatibility conditions. In Sec.\ \ref{sec3}, we analyze the Hamiltonian structure of the reduced system, classify its equilibria, and organize the phase space topology across different parameter regimes. We study a periodically forced extension of the reduced dynamics together with quantitative bifurcation and chaos diagnostics in order to investigate the transition from regular to irregular oscillatory behavor in Sec.~\ref{sec4}. In Sec.\ \ref{sec5}, we construct explicit traveling wave solutions and relate them to the discriminant structure of the Hamiltonian quadrature.  In Sec.\ \ref{sec6}, we validate the exact traveling wave branches by direct residual evaluation, numerical propagation of the extended nonlinear Helmholtz equation, and robustness tests under perturbed initial conditions. Finally, we summarize the main implications of the results and discuss the structural role of SS and SFS in nonparaxial nonlinear optical dynamics in Sec.\ \ref{sec7}.
	
	\section{Traveling wave reduction and dynamical system formulation}\label{sec2}
	\subsection{Symmetry guided traveling wave ansatz}
	We study the following NLH equation with self steepening (SS) and self frequency shift (SFS)
	\begin{equation}\label{eq1}
		\begin{split}
			i\frac{\partial \psi}{\partial x}
			+\frac{a_1}{2}\frac{\partial^2 \psi}{\partial t^2}
			+a_2 |\psi|^2 \psi
			+i \Big[a_3 \frac{\partial (|\psi|^2 \psi)}{\partial t}
			+a_4 \psi \frac{\partial (|\psi|^2)}{\partial t} \Big]
			+a_5 \frac{\partial^2 \psi}{\partial x^2}=0.
		\end{split}
	\end{equation}
	
	Here, $a_1$ is the second order dispersive coefficient, $a_2$ is the cubic Kerr coefficient, $a_3$ and $a_4$ describe non-Kerr nonlinear corrections from SS and SFS effects, and $a_5$ represents the nonparaxial correction. Our objective is to determine how the SS and SFS contributions modify the reduced phase space geometry and the admissible
	traveling wave states.
	
	Eq.~\eqref{eq1} is autonomous because it does not depend explicitly on $x$ or $t$. It is also invariant under the global phase transformation $\psi \mapsto \psi e^{i\theta}$. These translational and $U(1)$ gauge symmetries motivate the traveling wave reduction, where the envelope keeps a fixed shape in a moving frame and the carrier phase changes linearly. So, we introduce the following traveling wave ansatz
	\begin{align}\label{eq2}
		\psi(x,t)=V(\Xi)e^{i\xi},\qquad
		\Xi=x-\mu t,\qquad
		\xi=-\kappa x+\sigma t+\delta.
	\end{align}
	Here, $V(\Xi)\in\mathbb{R}$ is the real envelope amplitude, $\mu$ is the translation parameter, $\kappa$ is the carrier wavenumber, $\sigma$ is the carrier frequency, and $\delta$ is a constant phase. 
	
	Substituting Eq.\ \eqref{eq2} into Eq.\ \eqref{eq1}, and separating real and imaginary parts, we get
	
	\begin{align}
		\label{eq3}
		\Big(a_5+\frac{a_1}{2}\mu^2\Big)V^{''}
		+\Big(\kappa-\frac{a_1}{2}\sigma^2-a_5\kappa^2\Big)V
		+(a_2-a_3\sigma)V^3 &= 0, \\
		\label{eq4}
		(1-a_1\sigma\mu-2a_5\kappa)V'
		-\mu(3a_3+2a_4)V^{2}V' &= 0.
	\end{align}
	
	Eq.~\eqref{eq4} shows that a real traveling wave reduction only works if the carrier gradients, translation parameter, and higher order nonlinear coefficients are compatible.  For a nonconstant envelope with $V'\not\equiv 0$, Eq.~\eqref{eq4} must hold for all values of $V$. Therefore, the coefficients of
	$V'$ and $V^2V'$ must vanish independently, yielding
	\begin{align}\label{compatibility}
		1-a_1\sigma\mu-2a_5\kappa=0,
		\qquad
		3a_3+2a_4=0,
	\end{align}
	where the prime symbol indicates differentiation with respect to $\Xi$. These conditions show that the real-envelope traveling wave reduction is not valid for arbitrary non-Kerr nonlinear coefficients. Rather, the exact real traveling wave branches constructed below exist on the compatibility manifold defined by Eq.\ \eqref{compatibility}. So the imaginary part \eqref{eq4} is automatically satisfied and the problem reduces to the real second order envelope equation \eqref{eq3}. Equation\ \eqref{eq3} shows how different effects in the model stay balanced. The term with $V^{''}$ includes both dispersion and nonparaxial effects. The linear term explains how the carrier frequency and wavenumber interact. The cubic term shows that SS changes the strength of the nonlinearity through the factor $a_2 - a_3\sigma$. 

	\subsection{Planar dynamical system formulation}
	
	To characterize the local and global envelope dynamics induced by the traveling wave reduction, we rewrite the Eq.\ \eqref{eq3} as a first order planar system by introducing $Q(\Xi)=\frac{dV}{d\Xi}$, then Eq.~\eqref{eq3} takes the form
	\begin{align}\label{eq6}
		\frac{dV}{d\Xi} &= Q, \nonumber\\
		\frac{dQ}{d\Xi} &= \alpha V+\beta V^3,
	\end{align}
	where
	\begin{equation}\label{alphabeta}
		\alpha=\frac{a_1 \sigma^2+2a_5 \kappa^2-2\kappa}{2a_5+a_1 \mu^2},
		\qquad
		\beta=\frac{2(a_3 \sigma-a_2)}{2a_5+a_1 \mu^2}.
	\end{equation}
	
	\begin{figure}
		\centering
		\includegraphics[width=1.0\columnwidth]{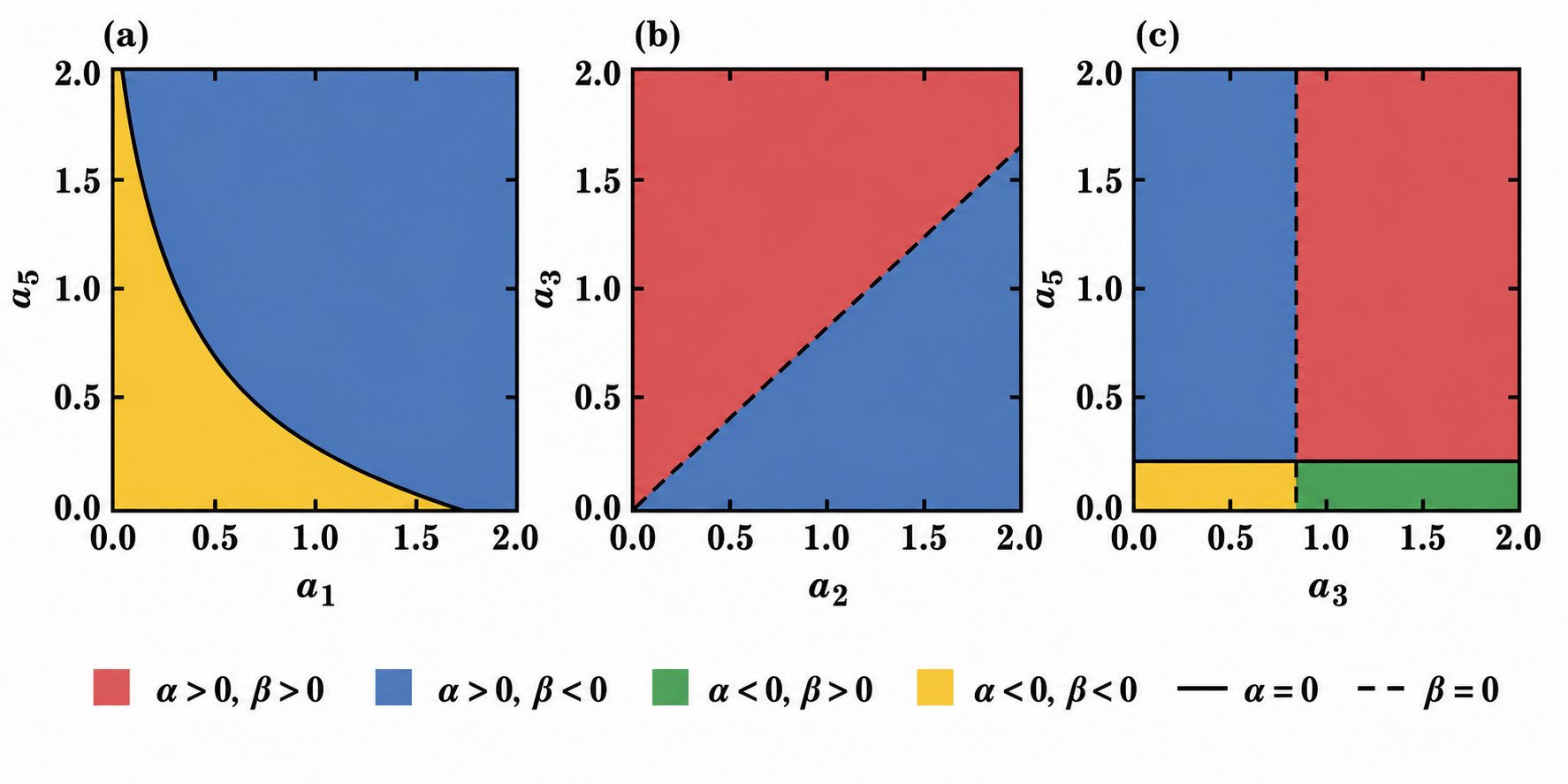}
		\caption{Parameter space diagrams illustrating the influence of the physical coefficients on the reduced Hamiltonian dynamics. (a) $(a_1,a_5)$-parameter plane with $a_2=1$, $a_3=0.5$, $a_4=-0.75$, $\mu=0.5$, $\sigma=1.2$, and $\kappa=\dfrac{1-a_1\sigma\mu}{2a_5}$. (b) $(a_2,a_3)$-parameter plane with $a_1=1$, $a_5=0.5$, $\mu=0.5$, $\sigma=1.2$, and $\kappa=0.4$ and $a_4=-3a_3/2$ imposed by the compatibility condition. (c) $(a_3,a_5)$-parameter plane with $a_1=1$, $a_2=1$, $\mu=0.5$, $\sigma=1.2$, and $\kappa=\dfrac{1-a_1\sigma\mu}{2a_5}$ and $a_4=-\dfrac{3a_3}{2}$ imposed by the compatibility condition. The diagrams demonstrate how different balances among dispersion, Kerr nonlinearity, self steepening, and nonparaxiality determine the signs of the effective coefficients $(\alpha,\beta)$ and consequently the corresponding phase space topology.}
		\label{fig0}
	\end{figure}
	
	Here, $\alpha$ denotes the effective linear restoring or destabilizing effect, and $\beta$ sets the sign and strength of the cubic nonlinear feedback.We explicitly relate the reduced coefficients $\alpha$ and $\beta$ to the original physical coefficients $a_{1}-a_{5}$ under the compatibility conditions given in Eq.~\eqref{compatibility}. Fig.\ \ref{fig0} shows that $\alpha$ and $\beta$ cannot be chosen independently or arbitrarily, because their admissible values are inherited from the balance among dispersion, Kerr nonlinearity, non-Kerr corrections, and nonparaxiality.
	
	Fig.~\ref{fig0}(a) presents the $(a_{1},a_{5})$-parameter plane for fixed $a_{2}=1$, $a_{3}=0.5$, $\mu=0.5$, and $\sigma=1.2$. The solid curve represents the condition $\alpha=0$ and separates the regions $\alpha>0,\beta<0$ and $\alpha<0,\beta<0$. Thus, for this parameter choice, the competition between the dispersion coefficient $a_{1}$ and the nonparaxial correction $a_{5}$ determines the sign of the effective linear coefficient $\alpha$, whereas the compatibility condition restricts the effective nonlinear coefficient to $\beta<0$.
	
	Fig.~\ref{fig0}(b) shows the $(a_{2},a_{3})$-parameter plane for fixed $a_{1}=1$, $a_{5}=0.5$, $\mu=0.5$, $\sigma=1.2$, and $\kappa=0.4$. The dashed line denotes the condition $\beta=0$ and separates the regimes $\alpha>0,\beta>0$ and $\alpha>0,\beta<0$. This panel demonstrates that the relative balance between the cubic Kerr coefficient $a_{2}$ and the SS coefficient $a_{3}$ controls the sign of the effective nonlinear coefficient $\beta$, while $\alpha$ remains positive. Therefore, even for the same effective linear response, variations in the original nonlinear coefficients may alter the cubic contribution and consequently modify the phase space structure.
	
	The simultaneous influence of non-Kerr and nonparaxial effects is illustrated in Fig.~\ref{fig0}(c), where $a_{3}$ and $a_{5}$ are varied with $a_{1}=a_{2}=1$, $\mu=0.5$, and $\sigma=1.2$. The vertical dashed line corresponds to $\beta=0$, whereas the horizontal solid line corresponds to $\alpha=0$. These boundaries divide the parameter space into the four sign combinations
	\[
	(\alpha>0,\beta>0), \qquad\\
	(\alpha>0,\beta<0), \qquad\\
	(\alpha<0,\beta>0), \qquad\\
	(\alpha<0,\beta<0).
	\]
	Each region is associated with a distinct reduced Hamiltonian configuration and, consequently, with different equilibrium arrangements, phase space topologies, and traveling wave families. Hence, the solutions of Eqs.\ \eqref{eq6}-\eqref{alphabeta} are not determined by unrestricted choices of $(\alpha,\beta)$. Instead, the accessible dynamical regimes are constrained by Eq.\ \eqref{compatibility} and arise directly from the physical balance among dispersion, cubic Kerr nonlinearity, self steepening, and nonparaxial effects in the original nonlinear Helmholtz equation.
	
	Equation~\eqref{eq6} is autonomous in $\Xi$ and polynomial in $V$, so its phase portrait is organised by the equilibrium points determined from $Q=0$ and $\alpha V+\beta V^3=0$, and by the stability type inferred from the linearization about each equilibrium. The signs and relative magnitudes of $\alpha$ and $\beta$ control the number of equilibria and the qualitative topology of trajectories, thereby providing a compact bifurcation parameterisation of traveling wave states. 
	
	This form is important for the analysis because it combines the effects of dispersion, nonparaxiality, carrier modulation, Kerr nonlinearity, and SS into the effective coefficients $\alpha$ and $\beta$. The SFS coefficient does not appear explicitly in the reduced coefficients. Instead, it enters through the compatibility condition $3a_3+2a_4=0$, which determines the admissible real traveling wave reduction. Consequently, the reduced dynamics is governed directly by $\alpha$ and $\beta$, while the SFS contribution restricts the parameter manifold on which the reduced Hamiltonian system is valid. From a physical viewpoint, Eq.~\eqref{eq6} shows that the SS coefficient $a_3$ directly modifies the reduced nonlinear dynamics through the effective cubic coefficient $\beta$, whereas the SFS coefficient $a_4$ acts indirectly through the compatibility condition $3a_3+2a_4=0$. Thus, SS governs the effective nonlinear response of the reduced Hamiltonian system, while SFS determines the admissible real traveling wave manifold without independently modifying the reduced coefficients $\alpha$ and $\beta$. Together, these non-Kerr effects influence the existence and organization of the traveling wave solutions.
	
	\section{Hamiltonian structure, equilibrium classification, and phase space organization}\label{sec3}
	
	\subsection{Hamiltonian structure and invariant phase plane geometry}
	
	The reduced system Eq.~\eqref{eq6} is autonomous and can be
	written equivalently as the second order nonlinear oscillator
	\begin{equation}\label{eq:oscillator}
		V^{''}=\alpha V+\beta V^3.
	\end{equation}
	This equation is autonomous and conservative, since the restoring term depends only on the variable $V$. It therefore admits a first integral, which provides the natural organizing principle for the phase plane geometry.
	
	To expose the conservative structure of the reduced system, we define the
	potential-like function
	\begin{equation}
		W(V)=\frac{\alpha V^{2}}{2}+\frac{\beta V^{4}}{4},
	\end{equation}
	such that
	\[
	W'(V)=\alpha V+\beta V^{3}.
	\]
	The corresponding conserved Hamiltonian is
	\begin{equation}
		H(V,Q)=\frac{Q^{2}}{2}-W(V)
		=\frac{Q^{2}}{2}
		-\frac{\alpha V^{2}}{2}
		-\frac{\beta V^{4}}{4}.
	\end{equation}
	Thus, the phase space geometry is governed by the combined structure of the kinetic term $\dfrac{Q^{2}}{2}$ and the Hamiltonian contribution
	\(-W(V)\). Depending on the signs of $\alpha$ and $\beta$, the Hamiltonian potential $-W(V)$ may exhibit a single-well, double-well, or inverted-well structure. Equivalently, the effective function $W(V)$ possesses the corresponding inverted geometry because the Hamiltonian is written in the form $H (V, Q)=\dfrac{Q^2}{2}-W(V)$. These configurations determine the number and type of equilibria, the existence of separatrices, and the families of bounded, localized, or transition-type traveling waves. 
	
	A direct computation verifies that \(\mathcal{H}\) is conserved along all solution trajectories. Indeed, using \(V'=Q\) and \(Q'=\alpha V+\beta V^3\), we find
	\begin{align*}
		\frac{d\mathcal{H}}{d\Xi}
		&=\frac{\partial \mathcal{H}}{\partial V}\frac{dV}{d\Xi}
		+\frac{\partial \mathcal{H}}{\partial Q}\frac{dQ}{d\Xi}\nonumber\\
		&=\left(-\alpha V-\beta V^3\right)Q
		+Q\left(\alpha V+\beta V^3\right)\nonumber\\
		&=0.
	\end{align*}
	Hence,
	\begin{equation*}\label{eq:solution_curve}
		\mathcal{H}(V(\Xi),Q(\Xi))=h,
	\end{equation*}
	where $h$ is constant along each solution curve.
	
	\par The conservation of $\mathcal{H}$ provides a direct geometric description of the dynamics of Eq.~\eqref{eq6}. The $(V,Q)$ phase plane is partitioned into invariant level sets $\mathcal{H}(V,Q)=h$, and trajectories cannot cross between different levels. As a result, the qualitative structure of the flow is determined by the topology of these constant-$\mathcal{H}$ curves. Equilibria occur at stationary points of $\mathcal{H}$ (equivalently, $Q=0$ and $W'(V)=0$), and their local character is governed by the linearization, in agreement with the local curvature of the Hamiltonian surface. In the remainder, we exploit this conserved quantity viewpoint to organise the phase plane description, i.e., separatrices, families of closed invariant curves (periodic motions) and solitary wave trajectories are identified as distinct energy levels $\mathcal{H}(V,Q)=h$ with different geometries in the $(V,Q)$ plane.
	These different shapes determine whether the reduced system admits one or three equilibria, whether those equilibria are elliptic or hyperbolic, and whether the corresponding wave states are oscillatory, localized, or transition-type. In this sense, the non-Kerr nonlinearity effects entering through the reduced coefficients $\alpha$ and $\beta$ act as structural modifiers of the effective energy landscape.

	\subsection{Equilibria and phase plane classification}\label{subsec:equilibria}
	The equilibria of Eq.\ \eqref{eq6} are $T_1=(0,0)$, and $T_{2,3}=\left(\pm\sqrt{-\frac{\alpha}{\beta}},\,0\right)$ whenever $\alpha\beta<0$. The existence and multiplicity of equilibria are therefore controlled entirely by the signs of $\alpha$ and $\beta$, providing a compact bifurcation parameterization of traveling wave states.\\
	The Jacobian matrix associated with Eq.\ \eqref{eq6} is
	\begin{equation}
		J(V,Q)=
		\begin{pmatrix}
			0 & 1\\
			\alpha+3\beta V^2 & 0
		\end{pmatrix}.
	\end{equation}
	For the trivial equilibrium $T_1=(0,0)$, the characteristic equation is $\lambda^2-\alpha=0$. Hence the eigenvalues are $\pm\sqrt{\alpha}$. Thus, if $\alpha>0$, $T_1$ is a saddle. If $\alpha<0$, the $T_1$ is a center in the linearized system, and also if $\alpha=0$, the equilibrium $T_1$ is nonhyperbolic.\\
	For the nontrivial equilibria $T_{2,3}=\left(\pm\sqrt{-\frac{\alpha}{\beta}},\,0\right)$, using $V^2=-\frac{\alpha}{\beta}$, we have $\alpha+3\beta V^2=\alpha-3\alpha=-2\alpha$.
	Therefore the corresponding eigenvalues satisfy $\lambda^2+2\alpha=0$.
	Hence, if $\alpha<0$, the nontrivial equilibria $T_{2,3}$ are saddles, and if $\alpha>0$, the nontrivial equilibria $T_{2,3}$, when they exist, are centers. 
	\begin{figure}
		\centering
		\includegraphics[width=1.0\columnwidth]{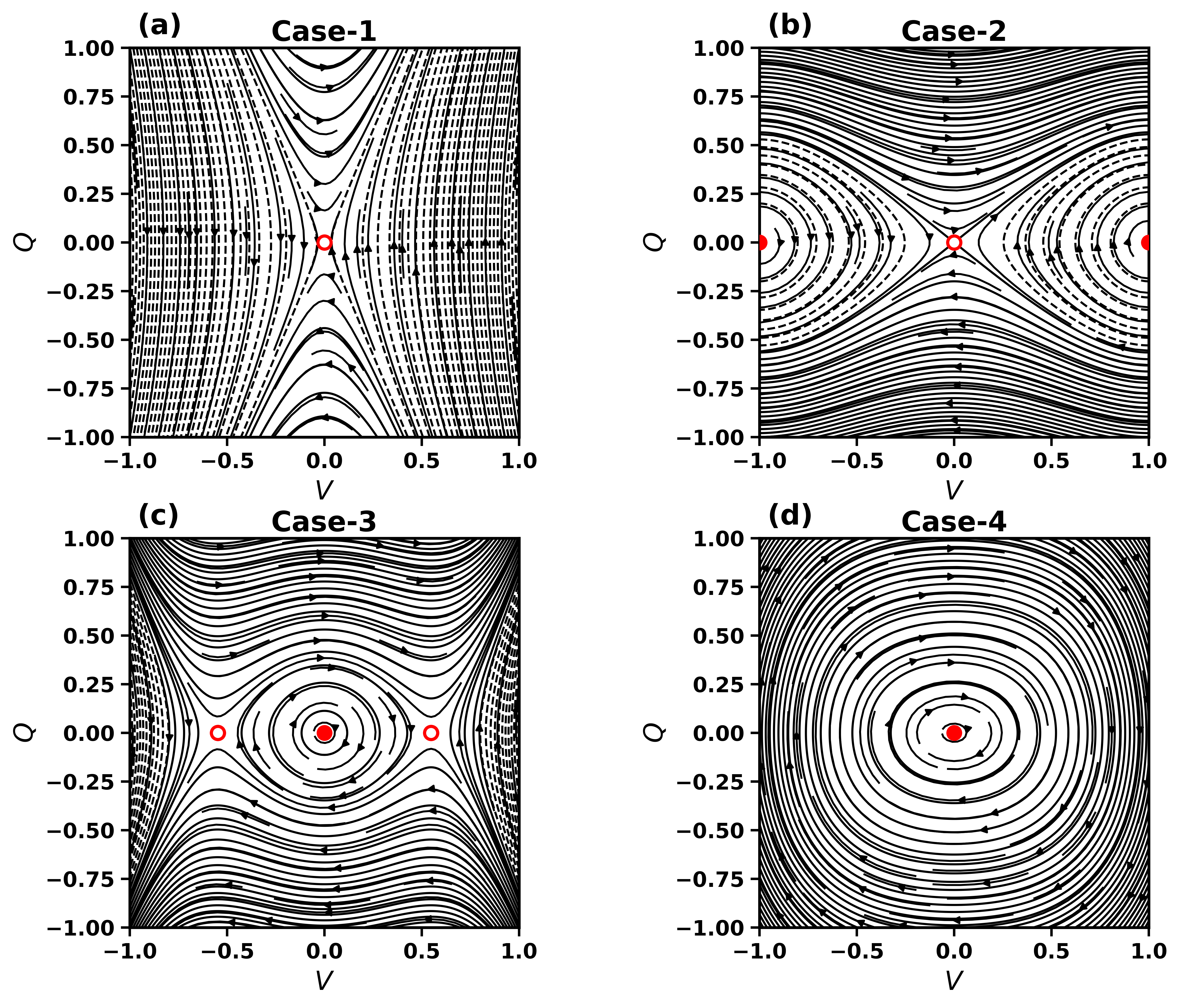}
		\caption{Representative phase portraits of the reduced planar system Eq.\ \eqref{eq6} for four sign combinations of \((\alpha,\beta)\). Each trajectory lies on an invariant Hamiltonian level set \(\mathcal{H}(V,Q)=h\). Saddles (red open circle) generate separatrices and organize global transitions between orbit classes, whereas centers (red filled circle) are surrounded by closed invariant curves corresponding to bounded oscillatory wave states. The representative reduced parameters are (a) \(\alpha=5.45\), \(\beta=0.4\); (b) \(\alpha=0.64\), \(\beta=-0.64\); (c) \(\alpha=-0.952\), \(\beta=3.17\); (d) \(\alpha=-0.55\), \(\beta=-1.2\).}
		\label{fig1}
	\end{figure}
	~~\\
	{\bf Case 1: $\alpha>0,~ \beta>0$}\\
	In this regime, $\alpha\beta>0$, only the trivial equilibrium $T_1$ exists since the nonzero equilibria $T_{2,3} $ become complex. The trivial equilibrium $T_1$ is a saddle as $\alpha>0$. The corresponding Hamiltonian potential $-W(V)$ does not possess a double-well structure in this parameter regime, so there are no additional potential wells supporting bounded oscillatory states away from the origin.
	Consequently, the phase portrait is dominated by the separatrix geometry associated with the single saddle. There are no bounded oscillatory islands centered away from the origin, and the traveling wave dynamics does not support the same type of localized bounded envelope states found in the mixed sign regimes.
	For the representative parameter set
	$a_1=a_2=1, ~a_3=0.5, ~a_4=-0.75, ~a_5=0.5, ~\mu=0.5, ~\sigma=2.5, ~\kappa=-0.25$,
	the reduced coefficients satisfy $\alpha>0,\beta>0$, and Fig.\ \ref{fig1}a displays the corresponding saddle dominated phase portrait.\\
	~~\\
	{\bf Case 2: $\alpha>0,~ \beta<0$}\\
	Here, $\alpha\beta<0$, and therefore three equilibria exist. The trivial equilibrium $T_1=(0,0)$ is a saddle because $\alpha>0$, whereas the two symmetric nonzero
	equilibria
	\[
	T_{2,3}=\left(\pm\sqrt{-\frac{\alpha}{\beta}},0\right)
	\]
	are centers. The reduced system consequently exhibits the canonical \textit{center-saddle-center} configuration. The conserved Hamiltonian level sets
	form two symmetric families of closed invariant curves around the nonzero centers, while the homoclinic separatrices associated with the saddle at the origin separate these bounded oscillatory regions. The closed trajectories surrounding $T_{2,3}$ correspond to periodic traveling wave solutions, whereas the homoclinic orbit to $T_1$ gives rise to a localized bright-type traveling wave. The presence of two
	symmetric centers further indicates two equivalent bounded envelope states within the reduced phase space geometry.
	
	For the representative parameter choice
	$a_1=a_2=1,~ a_3=0.5,~ a_4=-0.75,~ a_5=0.5,~
	\mu=0.5,~ \sigma=1.2,~ \kappa=0.4,$
	the compatibility conditions are satisfied and the reduced coefficients are
	\[
	\alpha=0.64,\qquad \beta=-0.64.
	\]
	Hence,
	\[
	T_{2,3}
	=
	\left(\pm\sqrt{-\frac{\alpha}{\beta}},0\right)
	=
	(\pm1,0),
	\]
	in agreement with the phase portrait shown in Fig.~\ref{fig1}(b).\\
	\\
	{\bf Case 3: $\alpha<0,~ \beta>0$}\\
	In this case $\alpha\beta<0$, so the three equilibria again exist. However, the stability arrangement is reversed. The $T_1$ is a center, while the nonzero equilibria $T_{2,3}$ are saddles as $\alpha<0$. The resulting geometry is a \textit{saddle--center--saddle} configuration. This regime is dynamically rich because the central elliptic region is bounded by separatrices attached to the two outer saddles. These separatrices define the transition between bounded oscillatory states around the origin and trajectories that connect distinct asymptotic branches. In the traveling wave interpretation, this structure is naturally associated with dark, front, or kink-type states, since the nonzero saddles provide the asymptotic anchors for transition profiles.\\
	For $a_1=a_2=1, ~a_3=0.5, ~a_4=-0.75, ~a_5=0.05, ~\mu=0.24, ~\sigma=2.5, ~\kappa=4.0$,
	the reduced coefficients satisfy $\alpha<0,~\beta>0$, and the phase portrait in Fig.~\ref{fig1}c indeed exhibits a center at the origin and symmetric saddle points near $(\pm 0.55, 0)$.\\
	\\
	{\bf Case 4: $\alpha<0,~ \beta<0$}\\
	In this regime, the origin is the only real equilibrium, and it
	is a center. The phase portrait therefore consists of closed invariant curves around \(T_1\), corresponding to bounded oscillatory states in the reduced envelope dynamics.Unlike Case 2, this regime has no additional nonzero equilibria. Unlike Case 3, it has no nontrivial saddles capable
	of generating separatrices. The dynamics is therefore topologically simpler, with oscillatory motion organized around a single elliptic core.\\
	For the representative parameter choice $a_1=a_2=1, ~a_3=0.5, ~a_4=-0.75, ~a_5=0.5, ~\mu=0.5, ~\sigma=0.5, ~\kappa=0.75$, 
	the system lies in the $(\alpha<0,\beta<0)$ regime, and Fig.~\ref{fig1}d shows the corresponding family of closed invariant curves around the origin.
	This classification immediately shows that the topology of the phase portrait is organized by the sign combination of $(\alpha,\beta)$. The four principal cases are summarized in Table\ \ref{Table_1}.
	\begin{table*}[t]
		\centering
		\small
		\renewcommand{\arraystretch}{1.2}
		\begin{tabular*}{\textwidth}{@{\extracolsep{\fill}}|c|c|c|l|l|l|}
			\hline
			\textbf{Case} & \textbf{$\alpha$} & \textbf{$\beta$} & \textbf{Phase space structure} & \textbf{Equilibrium configuration} & \textbf{Traveling wave interpretation} \\
			\hline
			
			1 & $>0$ & $>0$ &
			\parbox[t]{3.2cm}{Saddle-dominated geometry; no closed invariant curves} &
			\parbox[t]{4.2cm}{Only the trivial equilibrium $T_1=(0,0)$ exists, and it is a saddle; $T_{2,3}$ are not real} &
			\parbox[t]{6.4cm}{No bounded localized or oscillatory traveling wave states. The dynamics is governed by the separatrix of the single saddle, leading only to monotonic, separatrix-driven, or unbounded propagation states.} \\
			\hline
			
			2 & $>0$ & $<0$ &
			\parbox[t]{3.2cm}{Center--saddle--center configuration with a double-well structure} &
			\parbox[t]{4.2cm}{$T_1=(0,0)$ is a saddle, while the symmetric nonzero equilibria $T_{2,3}$ are centers} &
			\parbox[t]{6.4cm}{Supports bright-type localized states through homoclinic orbits to the central saddle, and periodic traveling waves through closed orbits around the two centers. This regime also reflects bistable bounded envelope organization.} \\
			\hline
			
			3 & $<0$ & $>0$ &
			\parbox[t]{3.2cm}{Saddle--center--saddle configuration; a central elliptic region is bounded by separatrices attached to the outer saddles} &
			\parbox[t]{4.2cm}{$T_1=(0,0)$ is a center, while the symmetric nonzero equilibria $T_{2,3}$ are saddles} &
			\parbox[t]{6.4cm}{Naturally supports dark solitons, fronts, and kink-type traveling waves. The nonzero saddles act as asymptotic anchor states for transition profiles, while the central closed orbits correspond to bounded oscillations around the origin.} \\
			\hline
			
			4 & $<0$ & $<0$ &
			\parbox[t]{3.2cm}{Single-center geometry with a family of closed invariant curves around the origin; no nontrivial saddles or separatrices} &
			\parbox[t]{4.2cm}{Only the trivial equilibrium $T_1=(0,0)$ exists, and it is a center} &
			\parbox[t]{6.4cm}{Supports only bounded oscillatory traveling wave states around the trivial background. No bright, dark, kink, or front-type solitary structures arise because homoclinic and heteroclinic connections are absent.} \\
			\hline
		\end{tabular*}
		\caption{Classification of the reduced phase space topology based on the sign combination of $(\alpha,\beta)$, together with the corresponding equilibrium configuration and traveling wave interpretation.}
		\label{Table_1}
	\end{table*}
	
	\section{Forced dynamics and routes to complex behavor}\label{sec4}
	
	\subsection{Periodically forced envelope system}\label{subsec:forced_system}
	We have investigated the reduced traveling wave dynamics derived from Eq.\ \eqref{eq1}. This equation is governed by the autonomous Hamiltonian system Eq.\ \eqref{eq6} in previous section. As shown earlier, this system is conservative and admits the Hamiltonian first integral
	$\mathcal{H}(V,Q)=h$, so that trajectories remain confined to invariant energy level sets in the $(V,Q)$ phase plane. In the traveling wave interpretation, this means that the envelope exchanges energy only internally between amplitude (shape of the wave) and 
	slope (how fast the shape changes), and the associated dynamics is organized by closed invariant curves, separatrices, or unbounded level sets depending on the signs of $\alpha$ and $\beta$.
	This conservative setting is mathematically fundamental, but it is not always sufficient for describing realistic propagation environments. In many physical systems, the medium is not perfectly isolated, and energy exchange arises from periodic pumping, refractive index modulation, nonlinear management, external excitation, and boundary forcing acting on the reduced envelope dynamics. In optical systems, such effects may arise through engineered photonic lattices, modulated waveguides, periodic refractive index structuring, or externally driven propagation conditions. In this context, the autonomous Hamiltonian oscillator \eqref{eq:oscillator} should be viewed as the intrinsic envelope dynamics, while externally imposed modulation acts as a perturbation that can break exact energy conservation and generate qualitatively new response regimes.
	
	For this purpose, we consider the periodically forced extension
	\begin{align}\label{MDS}
		\frac{dV}{d\Xi}&=Q, \nonumber\\
		\frac{dQ}{d\Xi}&=\alpha V+\beta V^3+\lambda\cos(\Gamma \Xi),
	\end{align}
	where $\lambda$ denotes the forcing amplitude, while $\Gamma$ represents the spatial wavenumber of the periodic modulation $\cos(\Gamma\Xi)$ along the traveling coordinate $\Xi$. Therefore, $\Gamma$ has units reciprocal to those of $\Xi$. In particular, if $\Xi$ represents a dimensional propagation distance measured in metres, then $\Gamma$ is measured in $\mathrm{m}^{-1}$. If $\Xi$ has been nondimensionalized, $\Gamma$ is dimensionless. Equivalently, the corresponding spatial modulation period is
	given by
	\[
	L_{\mathrm{mod}}=\frac{2\pi}{\Gamma}.
	\]
	The forcing term is introduced phenomenologically at the reduced system level to examine the dynamical consequences
	of periodic external modulation. It is not derived here from a specific modulation of the full NLH coefficients. Equation\ \eqref{MDS} can equivalently be written as
	\begin{equation}
		V''=\alpha V+\beta V^3+\lambda\cos(\Gamma\Xi).
		\label{eq:forced_oscillator}
	\end{equation}
	This equation represents a periodically forced Duffing-type nonlinear oscillator for the traveling wave envelope. The forcing term $\lambda\cos(\Gamma\Xi)$ introduces an externally imposed periodic modulation along the traveling coordinate while preserving the mathematical tractability of the reduced model. In contrast to the unforced Hamiltonian
	system, the forced system is nonautonomous and does not conserve the unperturbed Hamiltonian exactly. Instead, the periodic forcing produces successive energy input and extraction, which may reorganize the phase space structure and generate qualitatively distinct dynamical regimes.
	
	The introduction of forcing has several important consequences. First, the Hamiltonian \(\mathcal{H}\) is no longer conserved. Indeed, along the trajectories of Eq.\ \eqref{MDS},
	\begin{align}
		\frac{d\mathcal{H}}{d\Xi}
		&=
		\frac{\partial \mathcal{H}}{\partial V}\frac{dV}{d\Xi}
		+\frac{\partial \mathcal{H}}{\partial Q}\frac{dQ}{d\Xi}\nonumber\\
		&=
		(-\alpha V-\beta V^3)Q
		+Q\left(\alpha V+\beta V^3+\lambda\cos(\Gamma\Xi)\right)\nonumber\\
		&=
		\lambda Q\cos(\Gamma\Xi).
		\label{eq:Hdot_forced}
	\end{align}
	Thus, the sign of \(d\mathcal{H}/d\Xi\) changes dynamically, and the external drive may either inject or remove energy depending on the instantaneous phase relation between the oscillator velocity \(Q\) and the forcing \(\cos(\Gamma\Xi)\). Second, because the system is now explicitly non-autonomous, the dynamics is no longer restricted to invariant energy contours in the \((V,Q)\) plane. Third, the coexistence of an intrinsic nonlinear oscillation and an external forcing frequency allows resonant locking, multi-frequency response, torus-like motion, and irregular long-time trajectories.
	
\begin{figure}
		\centering
		\includegraphics[width=0.95\columnwidth]{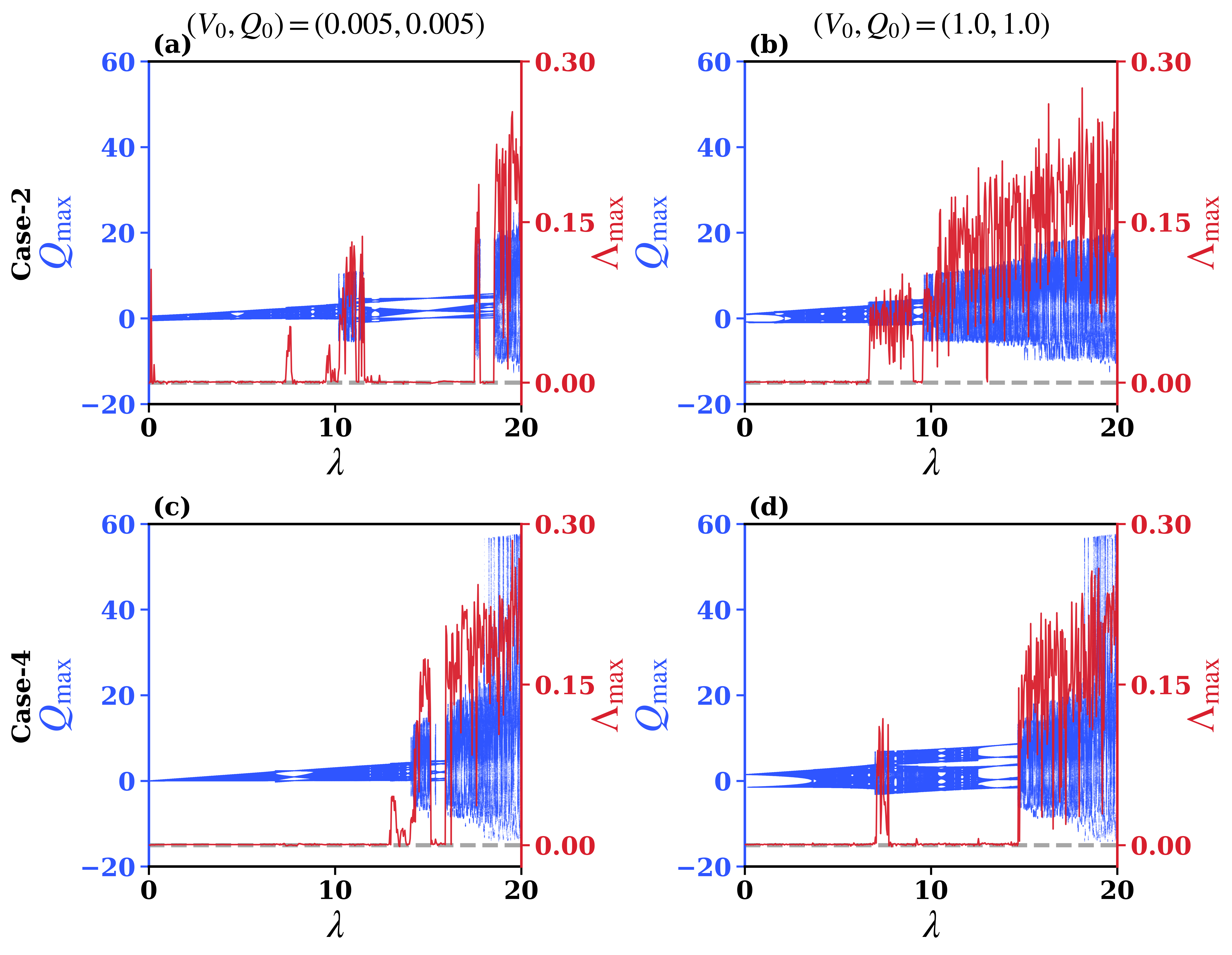}
		\caption{Bifurcation diagrams and largest Lyapunov exponent of the forced reduced system. Peak values \(Q_{\max}\) (blue) and largest Lyapunov exponent \(\Lambda_{\max}\) (red) are plotted as a function of forcing amplitude \(\lambda\) for two representative unforced regimes, Case 2 (\(\alpha=0.64,\beta=-0.64\), panels (a, b) and Case 4 (\(\alpha=-0.55,\beta=-1.2\), panels (c, d), with forcing frequency \(\Gamma=2\pi\). Panels (a, c) use the initial condition \((V_0,Q_0)=(0.005,0.005)\), while panels (b, d) use \((V_0,Q_0)=(1.0,1.0)\). The transition between regular quasiperiodic motion and irregular chaotic dynamics is identified in the bifurcation structure and confirmed by the largest Lyapunov exponent.}
		\label{fig2}
	\end{figure}

	To make this point more explicit, it is useful to rewrite the non-autonomous system Eq.\ \eqref{MDS} as an autonomous system in one higher dimension. Introducing the phase variable
	\begin{equation}
		\theta=\Gamma\Xi,\qquad \theta'=\Gamma,
	\end{equation}
	we obtain
	\begin{align}
		V' &= Q, \nonumber\\
		Q' &= \alpha V+\beta V^3+\lambda\cos\theta,\nonumber\\
		\theta' &= \Gamma.
		\label{eq:forced_autonomous}
	\end{align}
	The forced envelope dynamics therefore becomes a three dimensional autonomous flow. This is important mathematically because the unforced planar Hamiltonian system cannot exhibit chaos in the dissipative attractor sense because of its two-dimensional continuous
	flow and conserved first integral, whereas the forced three dimensional system can support substantially richer invariant sets and transition mechanisms. Here, forcing is introduced to address how external modulation reorganizes the traveling wave envelope dynamics of the nonlinear Helmholtz reduction and to determine the conditions under which regular oscillations evolve into increasingly complex responses. Once periodic forcing is introduced, the qualitative behavior of the envelope can no longer be inferred directly from the static Hamiltonian phase portraits of the unforced problem. The reduced dynamics may undergo transitions as the forcing amplitude \(\lambda\) increases, even when the underlying coefficients \(\alpha\) and \(\beta\) are held fixed. For this reason, a one-parameter bifurcation analysis in \(\lambda\) is exhibited in Fig.\ \ref{fig2}. The goal of this analysis is to determine how the asymptotic response of the forced envelope changes as the strength of external modulation is varied. It serves as a global summary of how the forced traveling wave envelope reorganizes across parameter space.
	
	\subsection{Bifurcation analysis of the forcing-induced transition between regular and irregular dynamics}
	To characterize the forcing induced transition structure, we construct a one-parameter bifurcation diagram by varying the forcing amplitude $\lambda$ in Eq.~\eqref{MDS} and collecting the local maxima of the momentum-like variable $Q(\Xi)=\frac{dV}{d\Xi}$. For each fixed $\lambda$, Eq.~\eqref{MDS} is integrated over a sufficiently long interval after discarding an initial transient, and the local maxima of \(Q(\Xi)\) $(Q_{max})$ over the post-transient regime, denoted the observation window by \([\Xi_{\mathrm{tr}},\Xi_{\mathrm{end}}]\)
	are plotted against $\lambda$. We focus on two representative unforced regimes identified earlier which are Case 2 $(\alpha=0.64, \beta=-0.64)$, corresponding to a center-saddle-center geometry in the unforced Hamiltonian system,
	and Case 4 $(\alpha=-0.55, \beta=-1.2)$, corresponding to a single-center geometry in the unforced Hamiltonian system. For each of these regimes, we consider two different initial conditions, \((V_0,Q_0)=(0.005,0.005)\) and \((V_0,Q_0)=(1,1)\), in order to assess whether the long-time forced response depends sensitively on the initial conditions within the underlying phase space structure. The resulting bifurcation diagrams are shown in Fig. \ref{fig2}. 
	
	\par For small forcing amplitude $\lambda$, the plotted $Q_{max}$ remains consistent with quasiperiodic motion, indicating a regular oscillatory response. As $\lambda$ increases, this structured response gives way to an irregular response, indicating the onset of chaos through the breakdown of quasiperiodicity at strong forcing \cite{ruelle1971nature, krysko2012routes} for both cases, Case 2 (Fig.\ \ref{fig2}(a)-(b)) and Case 4 (Fig.\ \ref{fig2}(c)-(d)). Although Case~2 and Case~4 remain physically and mathematically distinct because their unforced Hamiltonian phase portraits have different equilibrium structures, their forced responses display a comparable broad transition sequence as $\lambda$ increases. In both cases, weak forcing gives regular bounded quasiperiodic oscillation, and sufficiently strong forcing leads to irregular dynamics. This similarity refers only to the forcing-induced transition scenario and not to the detailed phase space geometry, amplitude range, or underlying equilibrium configuration.
	
	\par To provide a quantitative classification of the dynamics, the largest Lyapunov exponent $\Lambda_{\max}$ (described in APPENDIX \ref{appendix1}) is computed by varying forcing amplitude ($\lambda$) and is superimposed on the bifurcation diagrams in Fig.~\ref{fig2}. Values of $\Lambda_{\max}$ approaching zero are consistent with regular quasiperiodic motion, whereas a clearly positive exponent indicates sensitivity to initial conditions and supports the appearance of chaotic attractor. For both cases, the onset of positive $\Lambda_{\max}$ depends on the choice of initial condition. The correspondence between bifurcation-branch broadening and positive largest Lyapunov exponents identifies the irregular response as deterministic chaos emerging from a quasiperiodic state. As the forcing amplitude increases, the system exhibits alternating transitions between quasiperiodic and chaotic dynamics before settling into a persistently irregular chaotic regime.
	
	\begin{figure}
		\includegraphics[width=0.99\columnwidth]{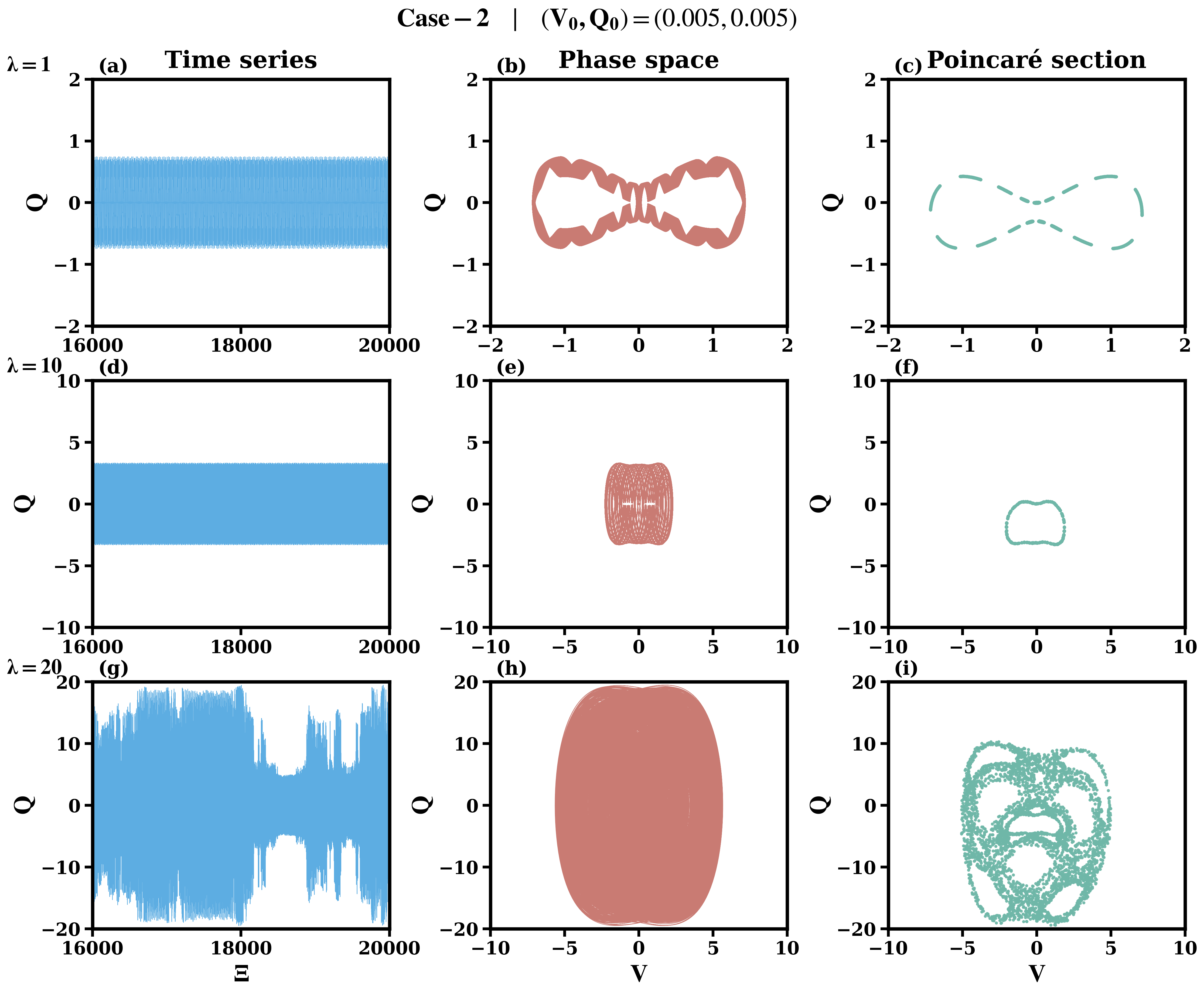}
		\caption{Panels (a), (d), and (g) show the time evolution of $Q(\Xi)$, panels (b), (e), and (h) present the corresponding phase space projections onto the $(V,Q)$ plane, and panels (c), (f), and (i) display the associated stroboscopic Poincar{\'e} sections. The upper, middle, and lower rows correspond to $\lambda=1$, $10$, and $20$, respectively, with $\alpha=0.64$, $\beta=-0.64$, and $\Gamma=2\pi$. The initial condition is $(V_0,Q_0)=(0.005,0.005)$. At low forcing amplitudes, the response remains bounded and smoothly modulated, with the phase space projections forming coherent torus-like bands. As the forcing strength increases, the oscillations become progressively irregular and the projected trajectories spread over a broader region of the $(V,Q)$ plane, indicating the breakdown of quasiperiodic organization and emergence of chaotic oscillation.}
		\label{fig3}
	\end{figure}

	\begin{figure}
		\includegraphics[width=0.99\columnwidth]{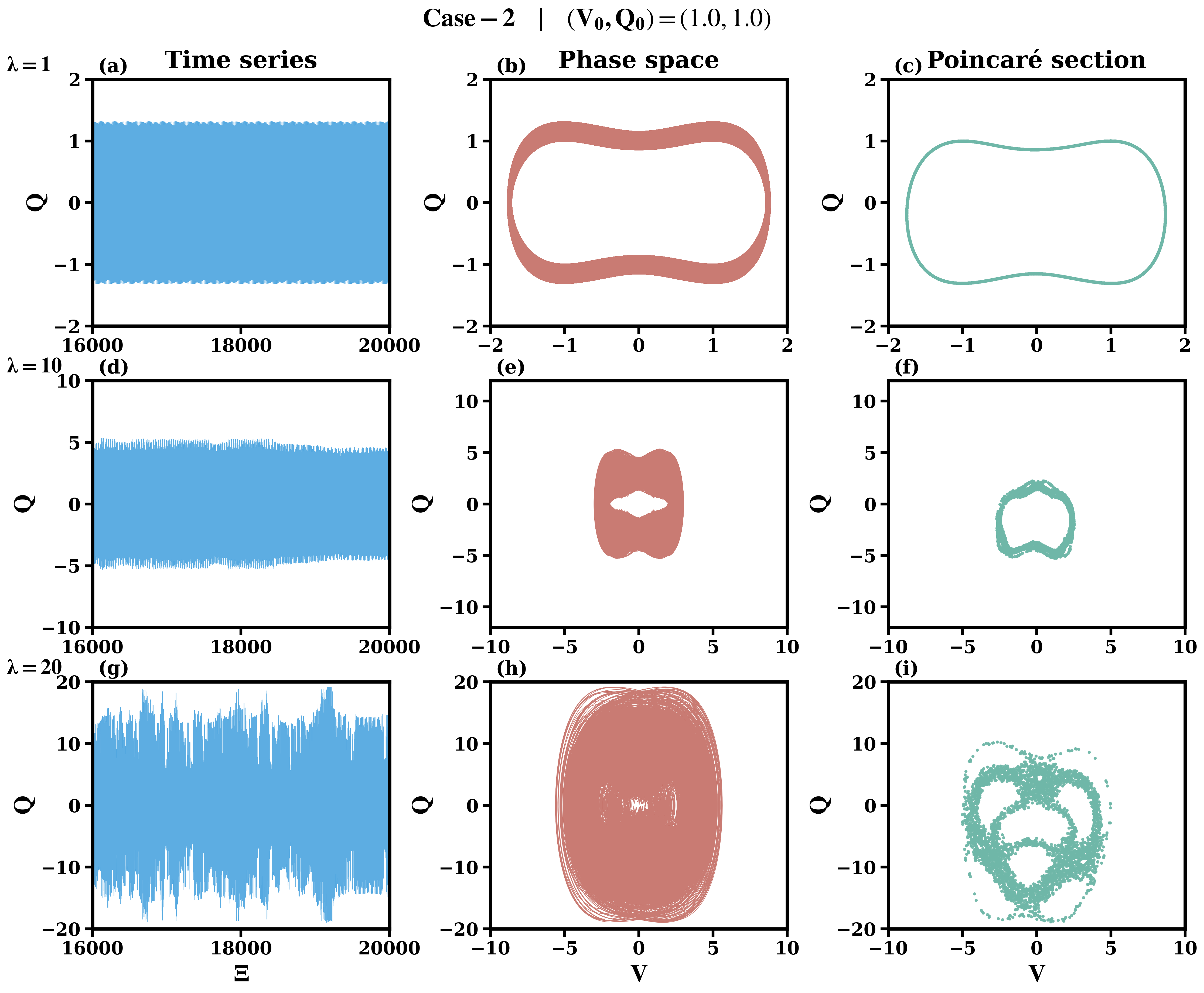}
		\caption{Panels (a), (d), and (g) show the time evolution of $Q(\Xi)$, panels (b), (e), and (h) present the corresponding phase space projections onto the $(V,Q)$ plane, and panels (c), (f), and (i) display the associated stroboscopic Poincar{\'e} sections. The upper, middle, and lower rows correspond to $\lambda=1$, $10$, and $20$, respectively, with \(\alpha=0.64\), \(\beta=-0.64\), and \(\Gamma=2\pi\). The initial condition is \((V_0,Q_0)=(1,1)\). The same qualitative progression from regular to increasingly irregular forced dynamics is observed, although the amplitude range and geometric spread differ from Fig.\ \ref{fig3}, reflecting the influence of the initial location relative to the underlying center--saddle--center structure of the unforced system.}
		\label{fig4}
	\end{figure}
	
	\begin{figure}
		\includegraphics[width=0.99\columnwidth]{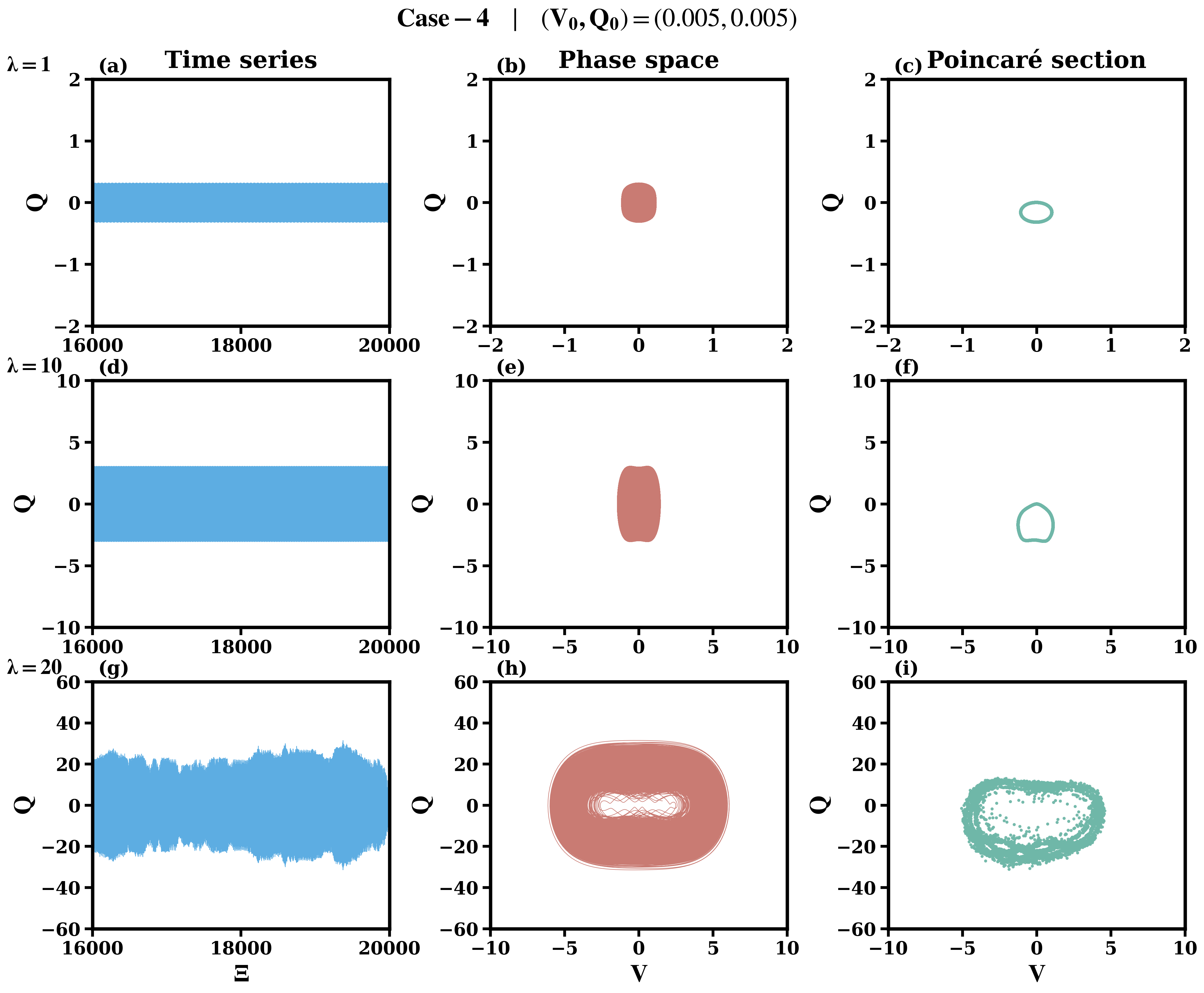}
		\caption{Panels (a), (d), and (g) show the time evolution of $Q(\Xi)$, panels (b), (e), and (h) present the corresponding phase space projections onto the $(V,Q)$ plane, and panels (c), (f), and (i) display the associated stroboscopic Poincar{\'e} sections. The upper, middle, and lower rows correspond to $\lambda=1$, $10$, and $20$, respectively, with \(\alpha=-0.55\), \(\beta=-1.2\), and \(\Gamma=2\pi\). The initial condition is \((V_0,Q_0)=(0.005,0.005)\). The low-forcing regime is organized around regular bounded oscillation near the single-center structure of the unforced system, while stronger forcing progressively broadens both the temporal response and the projected phase space geometry.}
		\label{fig5}
	\end{figure}
	
	\begin{figure}
		\includegraphics[width=0.99\columnwidth]{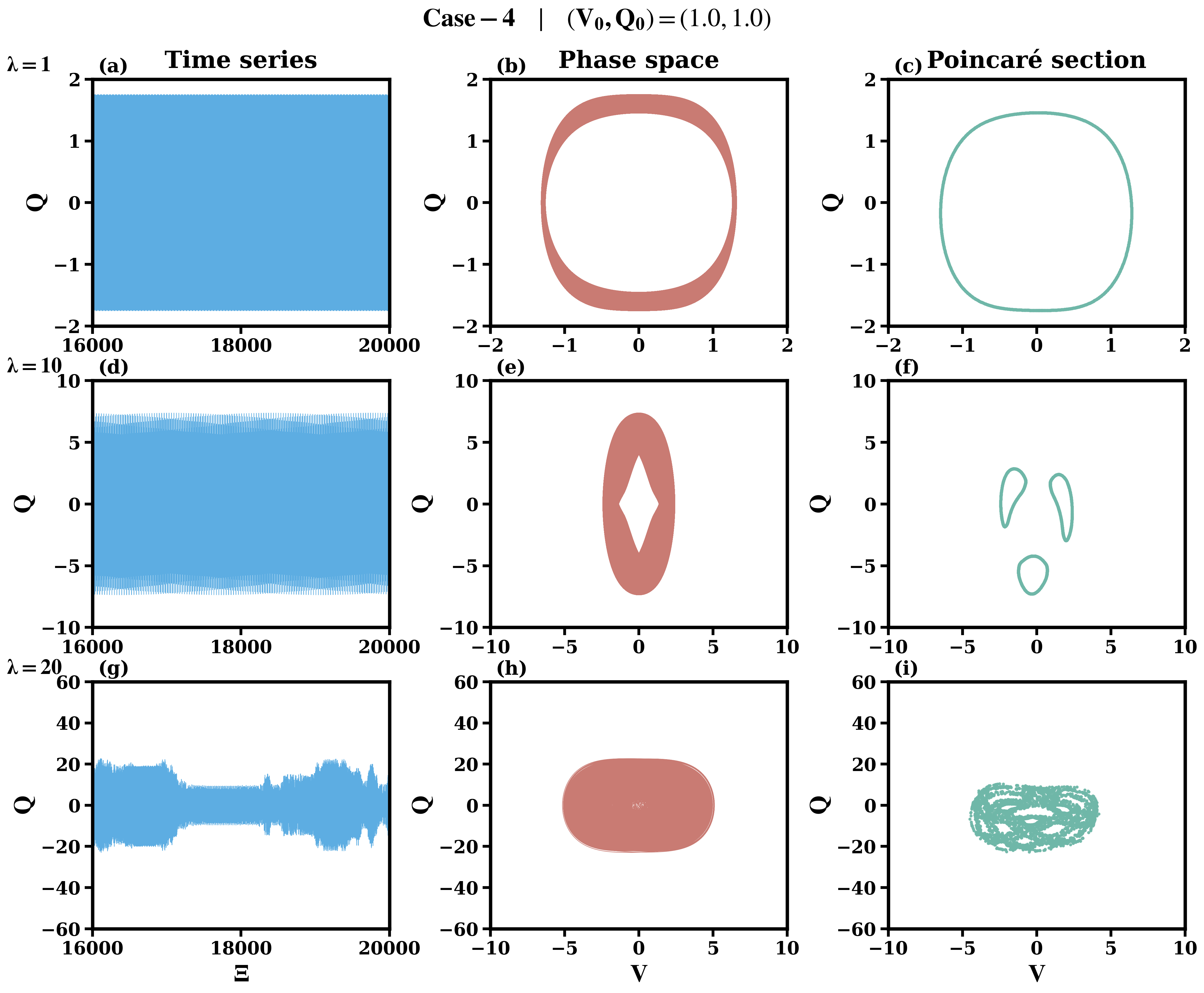}
		\caption{Panels (a), (d), and (g) show the time evolution of $Q(\Xi)$, panels (b), (e), and (h) present the corresponding phase space projections onto the $(V,Q)$ plane, and panels (c), (f), and (i) display the associated stroboscopic Poincar{\'e} sections. The upper, middle, and lower rows correspond to $\lambda=1$, $10$, and $20$, respectively, with \(\alpha=-0.55\), \(\beta=-1.2\), and \(\Gamma=2\pi\). The initial condition is \((V_0,Q_0)=(1,1)\). Although the same global transition from regular to irregular forced response is observed as in Fig.~\ref{fig5}, the trajectory amplitudes and phase plane spreading differ quantitatively, showing that the response retains sensitivity to the initial placement within the underlying Hamiltonian landscape.}
		\label{fig6}
	\end{figure}
	
	\subsection{Time series evolution, phase plane projections and Poincar{\'e} sections}
	We examine the temporal evolution of \( Q(\Xi) \) and the corresponding  \((V,Q)\) phase space and Poincar{\'e} section to characterize amplitude modulation, boundedness, and the emergence of different behavor of dynamics. These diagnostics provide a direct connection between the asymptotic structures observed in the bifurcation diagram (Fig.\ \ref{fig2}) and the underlying phase space dynamics. While the bifurcation diagram captures how the system varies with forcing, the time series, phase portraits, and Poincar{\'e} (described in APPENDIX \ref{appendix2}) section reveal the dynamical structure underlying these transitions.
	
	\begin{table}[t]
		\centering
		\begin{tabular}{ccccccc}
			\hline
			Case & $\alpha$ & $\beta$ & $(V_0,Q_0)$ & $\lambda$
			& $\Lambda_{\max}$ & Dynamical response \\
			\hline
			Case 2 & $0.64$ & $-0.64$ & $(0.005, 0.005)$
			& $1$  & $0.0004$ & Quasiperiodic \\
			&         &         &
			& $10$ & $0.0003$ & Quasiperiodic \\
			&         &         &
			& $20$ & $0.2130$ & Chaotic \\
			\hline
			Case 2 & $0.64$ & $-0.64$ & $(1.0, 1.0)$
			& $1$  & $0.0004$ & Quasiperiodic \\
			&        &         &
			& $10$ & $0.0712$ & Chaotic \\
			&        &         &
			& $20$ & $0.0852$ & Chaotic \\
			\hline
			Case 4 & $-0.55$ & $-1.20$ & $(0.005, 0.005)$
			& $1$  & $0.0004$ & Quasiperiodic \\
			&         &         &
			& $10$ & $0.0004$ & Quasiperiodic \\
			&         &         &
			& $20$ & $0.2275$ & Chaotic \\
			\hline
			Case 4 & $-0.55$ & $-1.20$ & $(1.0, 1.0)$
			& $1$  & $0.0003$ & Quasiperiodic \\
			&        &         &
			& $10$ & $0.0005$ & Quasiperiodic \\
			&        &         &
			& $20$ & $0.1843$ & Chaotic \\
			\hline
		\end{tabular}
		\caption{Largest Lyapunov exponents of the periodically forced reduced
			system for Cases 2 and 4, two initial conditions, and three forcing
			amplitudes. Values close to zero indicate
			quasiperiodic dynamics, whereas clearly positive values indicate
			chaotic dynamics.}
		\label{tab}
	\end{table}

	To elucidate the mechanism underlying the transition in Fig.~\ref{fig2}, we analyze \( Q(\Xi) \) and its \((V,Q)\) projection for forcing levels \( \lambda \in \{1,10,20\} \), with \( \Gamma = 2\pi \) (Figs.~\ref{fig3}--\ref{fig6}). Fig.~\ref{fig3} and \ref{fig4} correspond to Case~2 (\(\alpha = 0.64,\ \beta = -0.64\)), while Figs.~\ref{fig5} and \ref{fig6} correspond to Case~4 (\(\alpha = -0.55,\ \beta = -1.2\)). In each case, two initial conditions are considered, which are \((V_0,Q_0) = (0.005,0.005)\) and \((V_0,Q_0) = (1,1)\). For low to high forcing (\(\lambda = 1, 10, 20\)), the system exhibits bounded oscillatory behavior in both the time series of \(Q\) and the associated phase portraits.

	For Case~2 with the small initial condition $(V_0,Q_0)=(0.005,0.005)$ (Fig.~\ref{fig3}), the responses at $\lambda=1$ and $10$ remain bounded and exhibit organized amplitude modulation. The corresponding $(V,Q)$ projections form smooth or moderately thickened closed bands, while the stroboscopic Poincar{\'e} sections consist of closed invariant curves.
	Together, these signatures identify quasiperiodic motion on torus-like invariant sets in the extended $(V,Q,\theta)$ phase space. At $\lambda=20$, the temporal response becomes irregular, the phase space projection spreads over a substantially broader region, and the Poincaré section loses its smooth invariant-curve structure and develops an irregular scattered set. This qualitative change is supported by the increase of the largest Lyapunov exponent from values close to zero at $\lambda=1$ and $10$ to
	$\Lambda_{\max}=0.2130$ at $\lambda=20$ (Table~\ref{tab}), confirming the onset of chaotic dynamics.
	
	Similarly, the transition occurs for the same case but different choice of initial condition $(V_0,Q_0)=(1.0,1.0)$ (Fig.~\ref{fig4}). At $\lambda=1$, the time series remains regularly modulated, the phase space projection retains an organized banded geometry, and the Poincar{\'e} section is consistent with quasiperiodic motion, in agreement with $\Lambda_{\max}=0.0004$. By contrast, the responses at $\lambda=10$
	and $20$ display irregular temporal variability, and non-smooth Poincar{\'e} sets. The corresponding positive largest Lyapunov exponents, $\Lambda_{\max}=0.0712$ and $0.0852$, respectively, identify these states as chaotic. The results demonstrate that the initial condition strongly influences the forced response and can lead to different dynamical regimes at the same forcing amplitude ($\lambda$) within the underlying center-saddle-center phase space geometry.
	
	For Case~4, both initial conditions exhibit the same qualitative nature of the dynamics at a given forcing amplitude ($\lambda$), although the associated temporal and phase space structures differ markedly (Figs.~\ref{fig5} and \ref{fig6}). This interpretation is consistently supported by the time series, the $(V,Q)$ phase space projections, the Poincar{\'e} sections and the corresponding largest Lyapunov exponents reported in Table~\ref{tab}. The bifurcation diagram in Fig.~\ref{fig2}, the time series evolution, phase space projections and Poincar{\'e} sections
	in Figs.~\ref{fig3}--\ref{fig6}, and the largest Lyapunov exponents in Table~\ref{tab} provide mutually consistent evidence for a forcing-induced transition from quasiperiodic to chaotic dynamics. At low and moderate forcing, the dynamics is organized by smooth invariant curves in the stroboscopic section,
	consistent with motion on invariant tori. Increasing $\lambda$ deforms and ultimately destroys this organization, producing irregular Poincaré sets and positive largest Lyapunov exponents. These
	features are consistent with the destruction of invariant tori and a torus-breakdown transition to chaotic dynamics~\cite{ruelle1971nature,krysko2012routes,mishra2020routes}, although the precise transition threshold depends on the parameter regime and initial condition. The results should therefore be described as chaotic responses or chaotic invariant dynamics rather than dissipative chaotic attractors.
	
	\subsection{Effect of initial conditions and underlying Hamiltonian geometry}
	An additional observation emerging from Figs.~\ref{fig2}-\ref{fig6} is that the similar transition scenario occurs for both small amplitude and large amplitude initial conditions, but the detailed geometry of phase space of the response differs. The initial state determines where the forced trajectory begins relative to the unforced Hamiltonian structure, and therefore influences how readily the forcing can push the orbit across formerly invariant regions of phase space. In Case 2, where the unforced system contains a saddle and two nonzero centers, the response is more sensitive to whether the initial condition begins near the central separatrix structure or in one of the oscillatory wells. In Case 4, where the unforced system is organized around a single central oscillatory region, the distinction between initial conditions is weaker at small forcing but remains visible as the forcing grows. This comparison highlights that the forced dynamics does not erase the memory of the unforced Hamiltonian landscape immediately. Rather, the underlying geometry of the conservative traveling wave system continues to shape the response over a substantial range of forcing amplitudes. The emergence of irregular behavior under forcing must therefore be understood not only as an effect of external modulation, but also as a forcing induced reorganization of the intrinsic Hamiltonian phase space structure.

	\section{Exact traveling wave solutions and phase space consistency}
	\label{sec5}
	In this section, we construct exact traveling wave solutions of the
	unforced reduced envelope equation and organize them according to the
	Hamiltonian phase space structure described in Sec.\ \ref{sec3}. The reduced
	amplitude dynamics is governed by
	\begin{equation}
		V''=\alpha V+\beta V^3,
		\qquad
		\Xi=x-\mu t ,
		\label{eq:19}
	\end{equation}
	where the prime denotes differentiation with respect to the traveling
	coordinate $\Xi$. Multiplying Eq.~\eqref{eq:19} by $V'$ and
	integrating once gives
	\begin{equation}
		(V')^2
		=
		\alpha V^2+\frac{\beta}{2}V^4+2C_0
		\equiv P(V),
		\label{eq:secV_first_integral}
	\end{equation}
	where $C_0$ is an integration constant. Equivalently, the reduced system is
	Hamiltonian with
	\begin{equation}
		H(V,Q)
		=
		\frac{Q^2}{2}
		-
		\frac{\alpha V^2}{2}
		-
		\frac{\beta V^4}{4},
		\qquad Q=V',
		\label{eq:secV_Hamiltonian}
	\end{equation}
	which is conserved along every unforced trajectory. Hence, a real
	traveling wave solution exists only on intervals where
	\begin{equation}
		P(V)\geq 0.
		\label{eq:secV_real_condition}
	\end{equation}
	The admissible traveling wave families are determined not only
	by closed-form quadrature, but also by the compactness and topology of the corresponding Hamiltonian level set. This observation provides the direct connection between the exact solutions below and the phase portraits in
	Sec.\ \ref{sec3}.
	
	The full optical field is reconstructed from the real envelope using the traveling wave ansatz in Eq.\ \ref{eq2} which is
	\begin{equation}
		\psi(x,t)
		=
		V(x-\mu t)
		\exp\left\{\ii(-\kappa x+\sigma t+\delta)\right\}.
		\label{eq:secV_full_field}
	\end{equation}
	Here, $\mu$, $\kappa$, $\sigma$, and $\delta$ are the parameters appearing in
	the traveling wave ansatz. The reduced coefficients $\alpha$ and $\beta$ contain the
	combined contributions of dispersion, nonparaxiality, carrier
	modulation, Kerr nonlinearity, and SS. The SFS coefficient enters indirectly through the
	compatibility condition $3a_3+2a_4=0$.

	\subsection{Case 1: $\alpha>0,\ \beta>0$}
	\label{subsec:case_alpha_pos_beta_pos}
	
	For $\alpha>0$ and $\beta>0$, the origin is the only real equilibrium and
	it is a saddle. The Hamiltonian level sets are non-compact and no bounded
	oscillatory island exists. The quadrature admits the exact singular branch
	\begin{equation}
		V(\Xi)
		=
		\pm
		\sqrt{\frac{\alpha}{\beta}}\,
		\tan\left(
		\sqrt{\frac{\alpha}{2}}\,\Xi
		\right).
		\label{eq:secV_tan_branch}
	\end{equation}
	Therefore, the corresponding optical field is
	\begin{equation}
		\psi(x,t)
		=
		\pm
		\sqrt{\frac{\alpha}{\beta}}\,
		\tan\left[
		\sqrt{\frac{\alpha}{2}}\,(x-\mu t)
		\right]
		e^{\ii(-\kappa x+\sigma t+\delta)} .
		\label{eq:secV_psi_tan}
	\end{equation}
	Although Eq.~\eqref{eq:secV_psi_tan} is an exact solution of the reduced
	ordinary differential equation (ODE), it blows up at finite values of $\Xi$.
	It therefore corresponds to a non-compact Hamiltonian trajectory and should
	be regarded as an exact singular branch rather than a physically admissible
	bounded traveling wave.

	\subsection{Case 2: $\alpha>0,\ \beta<0$}
	\label{subsec:case_alpha_pos_beta_neg}
	
	For $\alpha>0$ and $\beta<0$, the origin is a saddle, while the two
	nontrivial equilibria
	\begin{equation}
		V_{\pm}
		=
		\pm
		\sqrt{-\frac{\alpha}{\beta}}
		\label{eq:secV_centers_caseII}
	\end{equation}
	are centers. This produces the center--saddle--center configuration
	identified in the Hamiltonian phase space analysis. This regime supports a
	bright homoclinic pulse attached to the saddle at the origin and closed
	periodic orbits around the two nonzero centers.

	
	The bounded periodic orbits surrounding the two nonzero centers can be
	written using the Jacobi elliptic function $\dn$ as
	\begin{equation}
		V(\Xi)
		=
		\pm A\,\dn(k\Xi,m),
		\qquad 0<m<1,
		\label{eq:secV_dn_solution}
	\end{equation}
	where
	\begin{equation}
		k^2=\frac{\alpha}{2-m},
		\qquad
		A^2=-\frac{2\alpha}{\beta(2-m)} .
		\label{eq:secV_dn_params}
	\end{equation}
	Therefore,
	\begin{equation}
		\psi(x,t)
		=
		\pm A\,\dn\left(k(x-\mu t),m\right)
		e^{\ii(-\kappa x+\sigma t+\delta)} .
		\label{eq:secV_psi_dn}
	\end{equation}
	For $m\to 0$, Eq.~\eqref{eq:secV_dn_solution} tends to the constant center
	state $V=\pm\sqrt{-\alpha/\beta}$.
	
	In the limit $m\rightarrow1$, the solution \eqref{eq:secV_psi_dn} becomes
	\begin{equation}
		\psi(x,t)
		=
		\pm
		\sqrt{-\frac{2\alpha}{\beta}}\,
		\sech\left[\sqrt{\alpha}\,(x-\mu t)\right]
		e^{\ii(-\kappa x+\sigma t+\delta)} .
		\label{eq:secV_psi_bright}
	\end{equation}
	This branch satisfies $V(\Xi)\to 0$ as $|\Xi|\to\infty$ and therefore
	corresponds to a homoclinic orbit to the saddle equilibrium at the origin.
	It is the bright-type localized traveling wave associated with the
	center-saddle-center phase space geometry.
	
	\subsection{Case 3: $\alpha<0,\ \beta>0$}
	\label{subsec:case_alpha_neg_beta_pos}
	
	For $\alpha<0$ and $\beta>0$, the origin is a center and the two nontrivial
	equilibria
	\begin{equation}
		V_{\pm}
		=
		\pm
		\sqrt{-\frac{\alpha}{\beta}}
		\label{eq:secV_saddles_caseIII}
	\end{equation}
	are saddles. The phase portrait contains a central family of closed orbits
	around the origin, together with separatrix connections involving the two
	outer saddles. This regime therefore supports front-type and bounded
	periodic traveling waves.
	
	The bounded closed orbits around the central equilibrium can be expressed
	using the Jacobi elliptic function $\sn$ as
	\begin{equation}
		V(\Xi)
		=
		A\,\sn(k\Xi,m),
		\qquad 0<m<1,
		\label{eq:secV_sn_solution}
	\end{equation}
	where
	\begin{equation}
		k^2=\frac{-\alpha}{1+m},
		\qquad
		A^2=\frac{-2m\alpha}{\beta(1+m)} .
		\label{eq:secV_sn_params}
	\end{equation}
	Thus,
	\begin{equation}
		\psi(x,t)
		=
		A\,\sn\left(k(x-\mu t),m\right)
		e^{\ii(-\kappa x+\sigma t+\delta)} .
		\label{eq:secV_psi_sn}
	\end{equation}
	For $0<m<1$, this branch gives a periodic oscillation about the center $V=0$.

	In the limit $m\rightarrow 1$, the solution \eqref{eq:secV_psi_sn} becomes
	\begin{equation}
		\psi(x,t)
		=
		\pm
		\sqrt{-\frac{\alpha}{\beta}}\,
		\tanh\left[
		\sqrt{-\frac{\alpha}{2}}\,(x-\mu t)
		\right]
		e^{\ii(-\kappa x+\sigma t+\delta)} .
		\label{eq:secV_psi_tanh}
	\end{equation}
	This solution satisfies
	\begin{equation}
		\begin{aligned}
			V(\Xi) &\to \mp\sqrt{-\frac{\alpha}{\beta}}
			\quad \text{as} \quad \Xi\to -\infty, \\
			V(\Xi) &\to \pm\sqrt{-\frac{\alpha}{\beta}}
			\quad \text{as} \quad \Xi\to +\infty .
		\end{aligned}
		\label{eq:secV_front_limits}
	\end{equation}
	It is therefore a heteroclinic connection between the two nonzero saddle
	equilibria and represents a transition-type, kink-type, or front-type
	traveling wave.
	
	\subsection{Case 4: $\alpha<0,\ \beta<0$}
	\label{subsec:case_alpha_neg_beta_neg}
	
	For $\alpha<0$ and $\beta<0$, the origin is the only real equilibrium and
	it is a center. There are no nontrivial saddle equilibria and therefore no
	homoclinic or heteroclinic separatrices. The admissible bounded solutions
	are only periodic oscillations around the origin.
	
	A convenient elliptic representation is
	\begin{equation}
		V(\Xi)
		=
		A\,\cn(k\Xi,m),
		\qquad
		0<m<\frac{1}{2},
		\label{eq:secV_cn_solution}
	\end{equation}
	with
	\begin{equation}
		k^2=\frac{-\alpha}{1-2m},
		\qquad
		A^2=
		\frac{2m(-\alpha)}{(-\beta)(1-2m)} .
		\label{eq:secV_cn_params}
	\end{equation}
	Therefore,
	\begin{equation}
		\psi(x,t)
		=
		A\,\cn\left(k(x-\mu t),m\right)
		e^{\ii(-\kappa x+\sigma t+\delta)} .
		\label{eq:secV_psi_cn}
	\end{equation}
	The restriction $0<m<1/2$ ensures $k^2>0$ and $A^2>0$. This family gives
	closed bounded Hamiltonian orbits around the single center. Since no saddle
	exists in this regime, neither bright homoclinic pulses nor front-type
	heteroclinic waves are supported.
	
	\subsection{Singular algebraic branches and admissibility}
	\label{subsec:singular_branches}
	The quartic quadrature also admits additional formal expressions involving $\tan$, $\coth$, $\sec$, and related functions. These branches are exact solutions of the reduced ordinary differential equation for suitable integration constants, but they generally exhibit finite-$\Xi$ singularities or lie on non-compact Hamiltonian level sets. Consequently, they should be distinguished from bounded traveling waves. In the phase space
	classification used here, physically admissible bounded branches are those associated with compact closed or separatrix level sets that remain finite for all $\Xi$. For this reason, the singular and non-compact formal branches are not analyzed and plotted because the analysis is restricted to bounded traveling wave profiles whose Hamiltonian trajectories remain regular and whose reconstructed intensity fields can be consistently validated over the full computational domain.
	The exact traveling wave branches, together with their corresponding phase space structures and physical interpretations, are summarized in Table\ \ref{Table_2}.
	
	\section{Validation of exact traveling wave branches}
	\label{sec6}
	In the previous sections, the extended nonlinear Helmholtz equation is
	reduced to a planar Hamiltonian system through the traveling wave
	transformation. The reduced dynamics provided a phase space classification of possible coherent structures, including localized, periodic, and transition-type traveling waves. In this section, we validate physical admissibility of exact traveling wave solutions of the reduced ordinary differential equation (ODE) with nonlinear Helmholtz equation (Eq.\ \eqref{eq1}). This step is necessary because a closed form solution of the reduced envelope equation is physically meaningful only if it also satisfies the proposed extended NLH equation under the required compatibility conditions. 
	
	\subsection{Residual validation of the analytical branches}
	
	The first validation step is based on direct substitution of each exact
	traveling wave branch into the full NLH equation. For this purpose, we define the residual operator
	\begin{equation}
		\begin{aligned}\label{eq:secV_residual}
			\mathcal{R}[\psi]
			&=
			i\psi_x
			+\frac{a_1}{2}\psi_{tt}
			+a_2|\psi|^2\psi  \\
			&\quad
			+i\left[
			a_3\frac{\partial(|\psi|^2\psi)}{\partial t}
			+a_4\psi\frac{\partial|\psi|^2}{\partial t}
			\right]
			+a_5\psi_{xx}.
		\end{aligned}
	\end{equation}
	A valid exact solution of the extended NLH equation must satisfy
	\begin{equation*}
		\mathcal{R}[\psi]\equiv 0.
		\label{eq:secV_residual_zero}
	\end{equation*}
	For numerical visualization, we plot
	$\log_{10}(|R[\psi]|+\epsilon)$, where $\epsilon=10^{-18}$ is introduced only as a logarithmic plotting offset to avoid the singularity of $\log_{10}(0)$. Residual values of this magnitude indicate excellent numerical agreement and provide strong evidence that the reconstructed traveling wave branch satisfies the full nonlinear Helmholtz equation, rather than only the reduced ordinary differential equation.
	
\begin{figure*}
		\centering
		\includegraphics[width=0.95\textwidth]{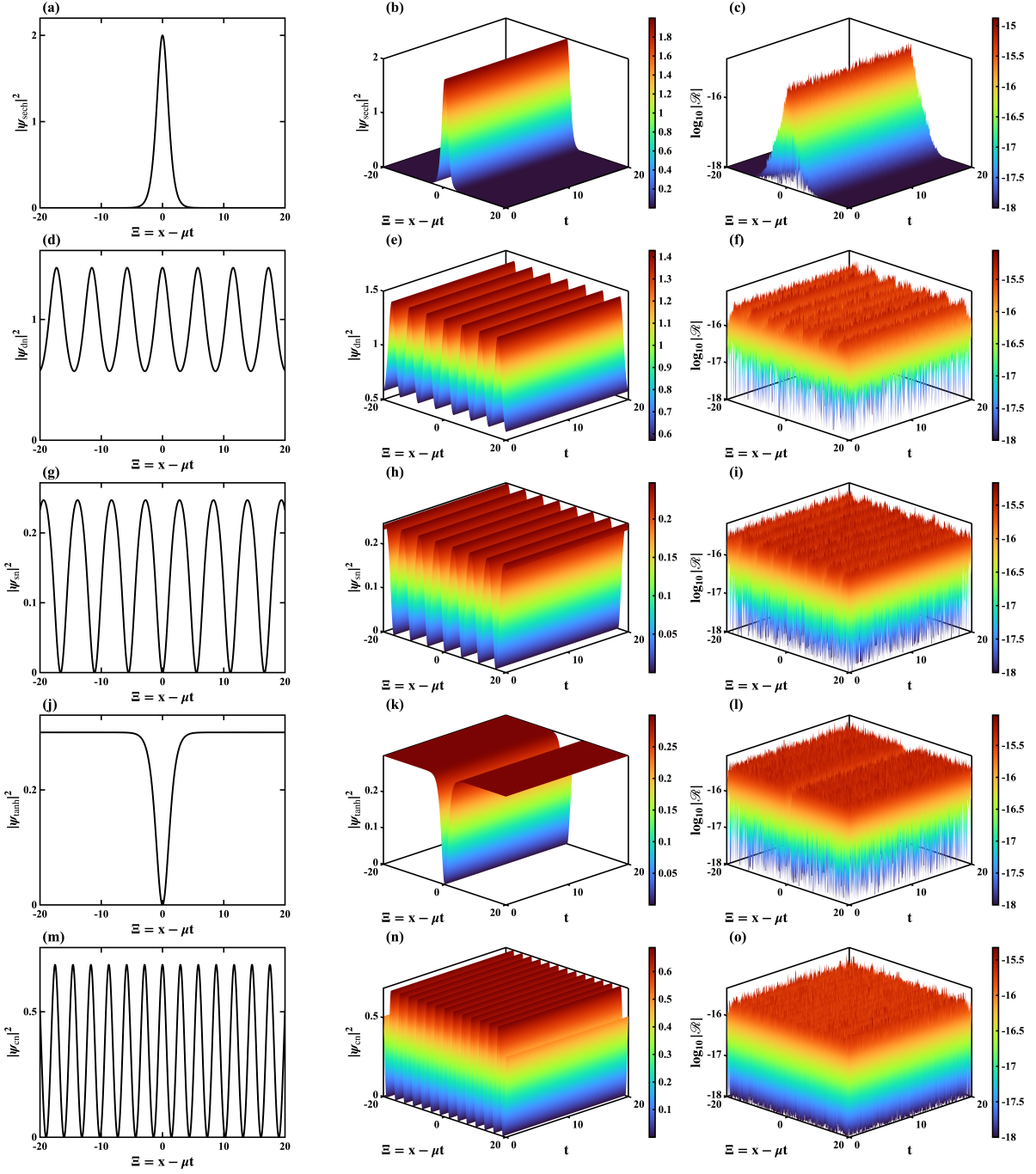}
		\caption{Analytical residual validation of the exact traveling wave branches against the full nonlinear Helmholtz equation. The rows correspond, respectively, to the localized bright $\sech$ branch in panels (a)-(c), the periodic $\dn$ branch in panels (d)-(f), the periodic $\sn$ branch in panels (g)-(i), the kink-type or dark-soliton $\tanh$ branch in panels (j)-(l), and the periodic $\cn$ branch in panels (m)-(o). The first column shows the one-dimensional intensity profile \(|\psi_i(\Xi)|^2=|V_i(\Xi)|^2\), where \(\Xi=x-\mu t\). The second column shows the corresponding spatiotemporal intensity surface \(|\psi_i(\Xi,t)|^2\). The third column shows the logarithmic full-NLH equation residual \(\log_{10}|\mathcal{R}|\), where \(\mathcal{R}\) is defined in Eq.~\eqref{eq:secV_residual}. The residual remains close to numerical round-off over the computational domain, confirming that the exact traveling wave branches satisfy the original extended nonlinear Helmholtz equation.}
		\label{fig7}
	\end{figure*}

	The five representative traveling wave branches considered in
	Fig.\ \ref{fig7} are
	\begin{align*}
		V_{\sech}(\Xi)
		&=
		\sqrt{-\frac{2\alpha}{\beta}}\,
		\sech(\sqrt{\alpha}\,\Xi),
		&& \alpha>0,\ \beta<0,
		\\
		V_{\dn}(\Xi)
		&=
		A_{\dn}\dn(k_{\dn}\Xi,m),
		&& \alpha>0,\ \beta<0,
		\\
		V_{\sn}(\Xi)
		&=
		A_{\sn}\sn(k_{\sn}\Xi,m),
		&& \alpha<0,\ \beta>0,
		\\
		V_{\tanh}(\Xi)
		&=
		\sqrt{-\frac{\alpha}{\beta}}\,
		\tanh\left(\sqrt{-\frac{\alpha}{2}}\,\Xi\right),
		&& \alpha<0,\ \beta>0,
		\\
		V_{\cn}(\Xi)
		&=
		A_{\cn}\cn(k_{\cn}\Xi,m),
		&& \alpha<0,\ \beta<0.
	\end{align*}
	The corresponding optical fields are reconstructed using the traveling wave ansatz in Eq.\ \ref{eq2} and subsequently substituted into Eq.\ \ref{eq1} for residual evaluation. Since the carrier factor has unit modulus,
	the physical intensity is
	\begin{equation*}
		|\psi_i(x,t)|^2
		=
		|V_i(x-\mu t)|^2,
		\qquad
		i\in\{\sech,\dn,\sn,\tanh,\cn\}.
		\label{eq:secV_intensity}
	\end{equation*}
	Therefore, the intensity structure is stationary in the moving coordinate
	\(\Xi=x-\mu t\), although the full optical field retains its carrier phase.

	Fig.~\ref{fig7} provides an analytical consistency check. The
	first column identifies the intensity morphology of each solution branch. The
	\(\sech\)-profile is a localized bright pulse and corresponds to a homoclinic structure in the reduced Hamiltonian phase plane. The \(\dn\)- and \(\cn\)-profiles represent bounded periodic wave trains. The \(\sn\)-profile is a periodic oscillatory branch arising in the saddle--center--saddle regime. The \(\tanh\)-profile is a transition-type branch connecting two nonzero asymptotic states and is therefore interpreted as a dark or kink-type traveling wave. The second column confirms that these intensity structures remain coherent in the \((\Xi,t)\)-representation. The third column is the most important validation component. The residual remains negligibly small, demonstrating that these profiles solve the full extended Helmholtz equation after reconstruction of the optical field.

	This residual validation is necessary for two reasons. First, the traveling wave reduction imposes compatibility conditions, without these conditions, a solution of the reduced ordinary differential equation (ODE) would not necessarily solve the full NLH equation. Second, the extended Helmholtz model contains non-Kerr effects and a longitudinal nonparaxial correction. Therefore, direct substitution into the full NLH equation, Eq.\ \eqref{eq1}, is required to confirm that the balance among
	dispersion, Kerr nonlinearity, self steepening, self frequency shift, and non paraxiality is preserved at the level of the original model.
	
	\subsection{Numerical Verification of traveling wave solutions}
	
	The residual calculation verifies the exact branches by direct substitution. To provide an independent computational check, we also solve the original NLH equation (Eq.\ \eqref{eq1})
	numerically. Equation\ \eqref{eq1} is written as a first order system by introducing
	\begin{equation*}
		u(x,t)=\psi(x,t),
		\qquad
		v(x,t)=\psi_x(x,t).
		\label{eq:secV_uv}
	\end{equation*}
	Then
	\begin{equation}
		u_x=v,
		\label{eq:secV_ux}
	\end{equation}
	and Eq.\ \eqref{eq1} gives
	\begin{equation}
		\begin{aligned}
			v_x
			=
			-\frac{1}{a_5}
			\bigg[
			&iv
			+\frac{a_1}{2}u_{tt}
			+a_2|u|^2u  \\
			&+
			i\left\{
			a_3\frac{\partial(|u|^2u)}{\partial t}
			+a_4u\frac{\partial |u|^2}{\partial t}
			\right\}
			\bigg].
		\end{aligned}
		\label{eq:secV_vx}
	\end{equation}
	Equations~\eqref{eq:secV_ux}--\eqref{eq:secV_vx} are integrated in \(x\)
	using a fourth-order Runge--Kutta scheme. The derivatives with respect to
	\(t\) are approximated using finite differences. The exact traveling wave
	solution is imposed at the initial plane \(x=x_0\). Accordingly, the computation should be interpreted as a
	controlled solver verification using manufactured exact
	boundary data rather than as an unconstrained propagation
	experiment. The numerical solution is then compared with the exact
	traveling wave profile at the final propagation plane \(x=x_{\mathrm{end}}\).

\begin{figure}
		\centering
		\includegraphics[width=0.99\columnwidth]{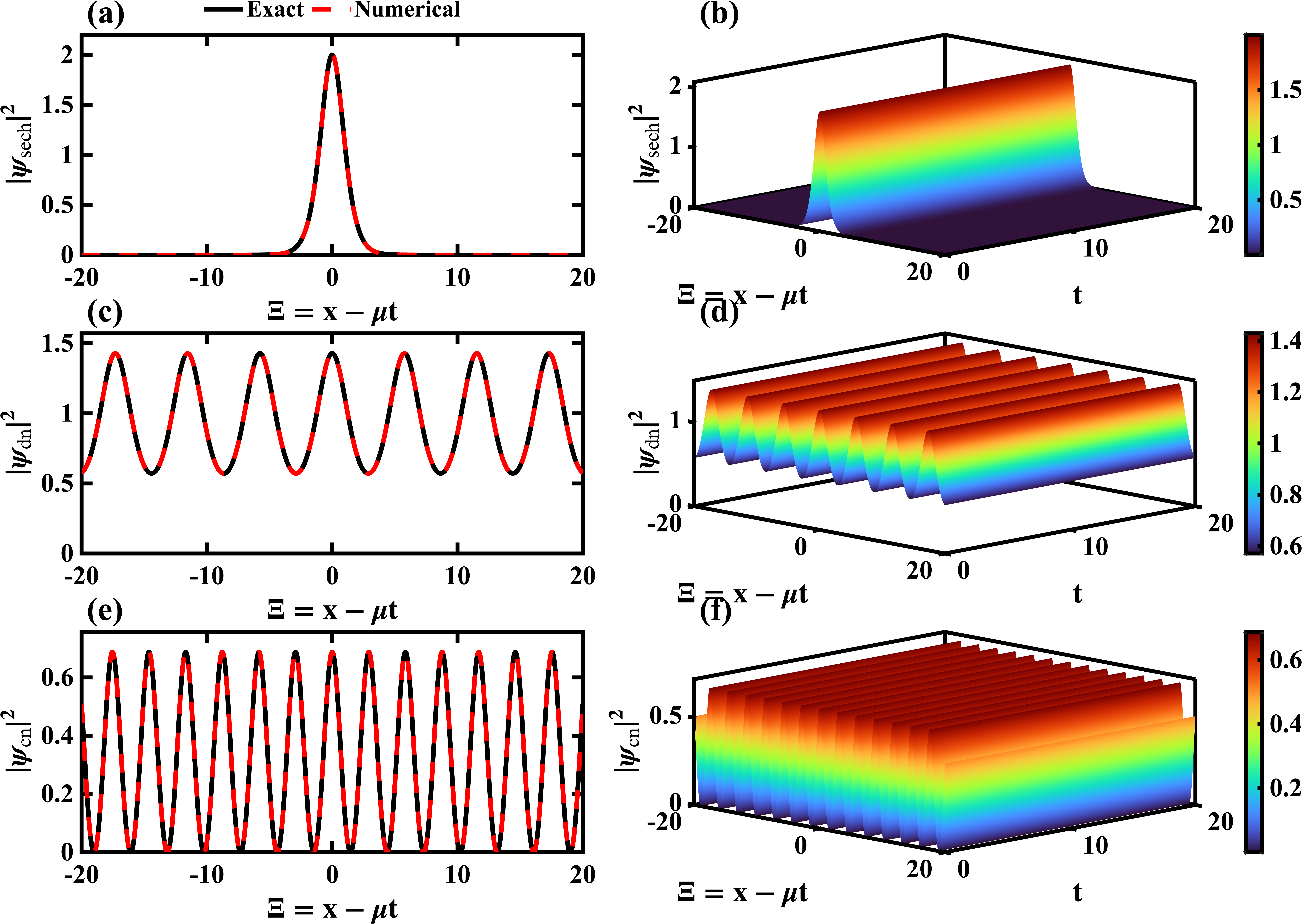}
		\caption{Panels (a) and (b) correspond to the localized bright $\sech$-type branch, panels (c) and (d) to the periodic $\dn$-type branch, and panels (e) and (f) to the periodic $\cn$-type branch. In the left column, panels (a), (c), and (e) compare the exact intensity profiles $|\psi_i|^2$ with the numerical profiles obtained after direct propagation of the full extended nonlinear Helmholtz equation. The exact solutions are shown by black solid curves, whereas the numerical solutions are represented by red dashed curves. In the right column, panels (b), (d), and (f) display the corresponding spatiotemporal intensity distributions $|\psi_i(\Xi,t)|^2$ over the selected propagation interval. The close agreement between the exact and numerically propagated profiles, together with the persistence of the associated spatiotemporal structures, confirms that the analytical traveling wave branches are accurately reproduced by the full-PDE solver.}
		\label{fig8}
	\end{figure}

	Figure\ \ref{fig8} presents the numerical validation of the selected \(\sech\), \(\dn\), and \(\cn\) branches, representing localized bright, periodic elliptic, and single-center periodic regimes, respectively. The black solid curves denote the exact analytical intensity profiles, while the red dashed curves denote the profiles obtained from direct numerical
	propagation. The close overlap between the two curves shows that the NLH equation solver preserves the traveling wave structure over the propagation interval. The right-column intensity surfaces further confirm that the coherent structure remains organized in the comoving coordinate \(\Xi=x-\mu t\). Fig.\ \ref{fig8} shows that the exact traveling wave solutions are dynamically reproducible under direct evolution
	of the full nonparaxial optical model. In other words, the solutions are not only algebraic profiles of the reduced ODE, they can also be propagated by the proposed NLH equation without losing their intensity structure over the selected computational window. This supports the claim that the balance among
	dispersion, Kerr nonlinearity, self steepening, self frequency shift, and the Helmholtz correction generates coherent optical traveling waves.
	
	\par Figure\ \ref{fig7} verifies all five analytical branches through full-equation residual evaluation, whereas Fig.\ \ref{fig8} demonstrates direct numerical reproducibility for the selected $\sech$, $\dn$, and $\cn$ branches of the NLH equation using partial differential equation (PDE) solver and comparing the numerical output with the exact profiles. These two validation steps close the logical gap between phase space analysis, exact solution construction, and PDE dynamics. The phase space analysis identifies which wave structures are admissible, the exact formulas give their closed form profiles, the residual calculation verifies that these profiles solve the full equation, and the direct numerical propagation confirms
	that the same structures are recovered by a full-PDE solver. Hence, Figs.~\ref{fig7} and \ref{fig8} establish that the
	localized, periodic, and transition-type branches are genuine coherent traveling wave states of the extended NLH model. This validation also clarifies the role of the non-Kerr nonlinear effects. The SS and SFS terms enter the reduced dynamics through the compatibility conditions and through the effective nonlinear coefficient \(\beta\). As a result, they do not merely perturb known solutions. Instead, they modify the admissible reduced phase space geometry and thereby control which traveling wave branches can exist. The agreement between analytical residual validation and direct numerical propagation therefore
	supports the main conclusion of the work, non-Kerr effects and nonparaxiality jointly reorganize the traveling wave landscape of the NLH system.
	
	\begin{table*}[t]
		\centering
		\small
		\renewcommand{\arraystretch}{1.2}
		\begin{tabular*}{\textwidth}{@{\extracolsep{\fill}}|c|l|l|l|}
			\hline
			\textbf{Parameter regime}  & \textbf{Phase space structure} & \textbf{Admissible solutions} & \textbf{Singular branches} \\
			\hline
			$\alpha>0,\ \beta>0$
			&
			\parbox[t]{3.2cm}{Single saddle}
			&
			\parbox[t]{3.2cm}{No bounded traveling wave ~~~~~~~~~~~~~~~~~~~~~~~~~~~~~~~~~~~~~~~~~~~~~~~~~~~~~~}
			&
			\parbox[t]{3.0cm}{singular trigonometric branches, including tangent-type solutions}
			\\
			\hline
			
			$\alpha>0,\ \beta<0$
			&
			\parbox[t]{3.2cm}{Center--saddle--center}
			&
			\parbox[t]{3.2cm}{Bright homoclinic $\sech$ branch, periodic $\dn$ branch around the two nonzero centers ~~~~~~~~~~~~~~~~~~~~~~~~~~~~~~~~~~~~~~~~~~~~~~~~~~~~~~~}
			&
			\parbox[t]{3.0cm}{Singular hyperbolic branches}
			\\
			\hline
			
			$\alpha<0,\ \beta>0$
			&
			\parbox[t]{3.2cm}{Saddle--center--saddle}
			&
			\parbox[t]{3.2cm}{Kink $\tanh$ branch, periodic $\sn$ branch around the origin ~~~~~~~~~~~~~~~~~~~~~~}
			&
			\parbox[t]{3.0cm}{Singular $\coth$-type branches}
			\\
			\hline
			
			$\alpha<0,\ \beta<0$
			&
			\parbox[t]{3.2cm}{Single center}
			&
			\parbox[t]{3.2cm}{Periodic $\cn$ branch around the origin ~~~~~~~~~~~~~~~~~~~~~~~~~~~~~~~~~~~~~~~~~~~}
			&
			\parbox[t]{3.0cm}{No regular separatrix branch}
			\\
			\hline
			
		\end{tabular*}
		\caption{Phase space consistent classification of exact traveling wave solutions of
			the reduced envelope equation $V''=\alpha V+\beta V^3$. Here $\Xi=x-\mu t$, and the full optical field is $\psi(x,t)=V(\Xi)e^{\ii(-\kappa x+\sigma t+\delta)}$. The constant $C_0$ denotes the integration constant in $(V')^2=\alpha V^2+\frac{\beta}{2}V^4+2C_0$.}
		\label{Table_2}
	\end{table*}

	\begin{figure*}
		\centering
		\includegraphics[width=0.99\textwidth]{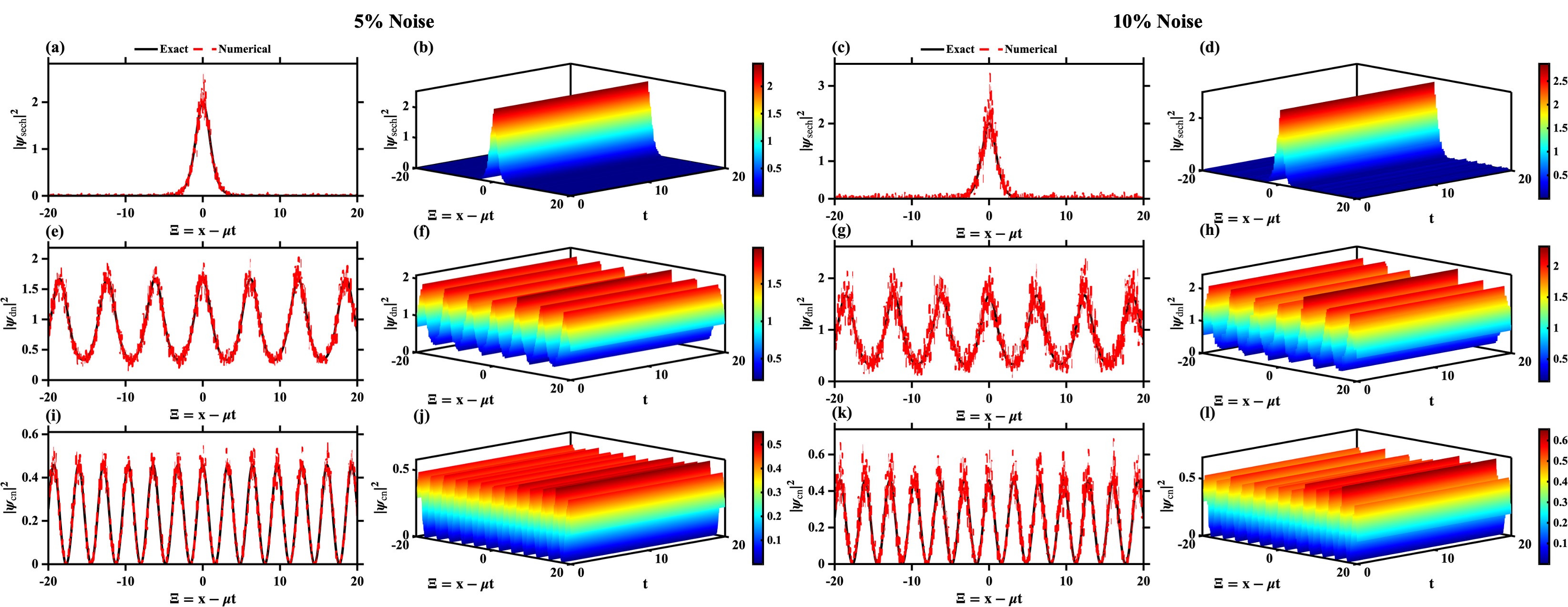}
		\caption{Numerical robustness of selected traveling wave solutions under perturbed initial conditions. Comparison of the analytical and numerically propagated intensity profiles of the nonlinear Helmholtz equation under $5\%$ and $10\%$ complex Gaussian white-noise perturbations. The upper, middle, and lower rows correspond to the sech-type bright solution, the $\mathrm{dn}$-type periodic solution with $m=0.80$, and the $\mathrm{cn}$-type periodic solution with $m=0.25$, respectively. Panels (a), (e), and (i), and panels (c), (g), and (k), compare the exact intensities, shown by black solid curves, with the corresponding numerical solutions, shown by red dashed curves, at the final propagation distance for $5\%$ and $10\%$ noise, respectively. Panels (b), (f), and (j), and panels (d), (h), and (l), show the corresponding numerical intensity distributions in the traveling coordinate $\Xi=x-\mu t$ for $5\%$ and $10\%$ perturbations, respectively. The persistence of the localized and periodic profiles, accompanied only by bounded noise-induced fluctuations, demonstrates the robustness of all three tested traveling wave branches against small perturbations of the initial field. The noise amplitude is defined relative to the peak amplitude of each exact solution.}
		\label{fig9}
	\end{figure*}

	\subsection{Numerical robustness under perturbed initial conditions}
	
	The analytical residual validation presented in the previous subsection confirms that the obtained traveling wave branches satisfy the full nonlinear Helmholtz equation. However, analytical consistency alone does not guarantee their physical relevance. In realistic nonlinear optical media, propagating optical pulses are inevitably affected by fabrication imperfections, environmental fluctuations, and input noise. Therefore, it is important to examine whether the exact traveling wave solutions remain dynamically robust under small perturbations. In the present work, this robustness is investigated by perturbing the exact traveling wave profile with additive discrete complex Gaussian perturbations and subsequently propagating the perturbed field using the full nonlinear Helmholtz equation. During propagation, the numerical solution is continuously compared with the corresponding analytical traveling wave branch. Throughout this study, a traveling wave solution is regarded as dynamically robust if it preserves its localization or periodicity, overall waveform, amplitude, and propagation characteristics while the perturbation remains bounded without causing qualitative waveform deformation or propagation instability.
	
	To examine the robustness of the traveling wave solutions, the perturbed initial condition is prescribed as
	
	\begin{equation}
		\psi_{\mathrm{pert}}(x_0,t)
		=
		\psi_{\mathrm{exact}}(x_0,t)
		+
		\frac{\delta A_{\mathrm{ref}}}{\sqrt{2}}
		\left[\eta_1(t)+i\eta_2(t)\right],
	\end{equation}
	
	where
	$A_{\mathrm{ref}}
	=
	\max_t
	\left|
	\psi_{\mathrm{exact}}(x_0,t)
	\right|
	$
	denotes the peak amplitude of the corresponding exact traveling wave solution. The random variables $\eta_1(t)$ and $\eta_2(t)$ are independent standard Gaussian random variables sampled at the discrete temporal grid points $t_j$ with zero mean and unit variance representing the real and imaginary components of the perturbation. The factor $\dfrac{1}{\sqrt{2}}$ normalizes the complex perturbation to unit mean-square magnitude, while $\delta$ denotes the relative perturbation strength. In the present simulations, $\delta=0.05$ and $\delta=0.10$ are employed to represent $5\%$ and $10\%$ complex Gaussian perturbations, respectively.
	
	Since the nonlinear Helmholtz equation is second order with respect to the propagation coordinate $x$, the numerical formulation introduces the auxiliary variable $v=\psi_x$, thereby converting the governing equation into an equivalent first-order system. Accordingly, both $\psi(x_0,t)$ and $\psi_x(x_0,t)$ must be specified at the initial propagation plane. In the present computations, only the optical field is perturbed according to the above equation, whereas the corresponding propagation derivative is initialized from the exact traveling wave solution,
	\[
	v(x_0,t)
	=
	\left.
	\frac{\partial\psi_{\mathrm{exact}}(x,t)}
	{\partial x}
	\right|_{x=x_0}.
	\]
	This initialization isolates the influence of random perturbations in the optical field while preserving the exact propagation derivative required by the first-order formulation. The resulting initial data are then propagated using the same fourth-order Runge--Kutta finite-difference scheme employed in the previous validation subsection.
	
	Figure~\ref{fig9} presents the numerical stability of three representative traveling wave branches under both $5\%$ and $10\%$ complex Gaussian perturbations. The first row corresponds to the localized bright $\mathrm{sech}$ solution, the second row to the periodic $\mathrm{dn}$ solution, and the third row to the periodic $\mathrm{cn}$ solution. For each traveling wave branch, the first and third columns [panels (a), (e), (i) and (c), (g), (k)] compare the analytical intensity profile (black solid curve) with the numerically propagated solution (red dashed curve) at the final propagation distance for $5\%$ and $10\%$ perturbations, respectively. The second and fourth columns [panels (b), (f), (j) and (d), (h), (l)] display the corresponding spatiotemporal intensity evolution.
	
	For the localized $\mathrm{sech}$ branch, the propagated numerical solution remains in excellent agreement with the analytical profile under both perturbation levels, indicating that the pulse preserves its amplitude, width, and localization throughout the propagation interval. The periodic $\mathrm{dn}$ and $\mathrm{cn}$ traveling wave branches likewise maintain their periodic structures, with only small bounded oscillatory deviations from the analytical solutions. Increasing the perturbation level from $5\%$ to $10\%$ produces slightly larger fluctuations but does not alter the qualitative propagation behavor or destroy the underlying traveling wave structures.
	
	Figure\ \ref{fig9} demonstrates that all representative traveling wave branches retain their characteristic localized or periodic profiles in the presence of moderate random perturbations. The numerical solutions remain in close agreement with the corresponding analytical traveling wave branches throughout the computational domain, and no qualitative waveform breakdown is observed over the selected computational interval. These numerical experiments therefore complement the residual validation presented in the previous subsection and demonstrate that the obtained traveling wave solutions are not only exact solutions of the reduced Hamiltonian system but also numerically robust over the selected propagation interval and for the tested perturbation levels.
	\section{Conclusion}\label{sec7}
	\par In this work, we have investigated the nonlinear Helmholtz equation incorporating self steepening and self frequency shift, with the aim of clarifying how non-Kerr nonlinear effects reshape wave dynamics beyond the paraxial regime. This problem is relevant to the propagation of ultrashort and high-intensity optical pulses, where nonparaxiality and non-Kerr nonlinearity can play a decisive role in determining waveform structure, stability, and propagation characteristics. Against this background, the present study addresses a gap between exact solution approaches and nonlinear dynamical analysis by developing a unified framework that combines traveling wave reduction, Hamiltonian dynamical systems theory, phase space analysis, bifurcation analysis, exact traveling wave construction, and numerical validation within the same nonlinear Helmholtz model.
	
	\par Using a symmetry-based traveling wave ansatz, we reduced the proposed NLH equation to a planar dynamical system and showed that its qualitative behavior is organized by the reduced parameter structure. This formulation enabled a systematic classification of equilibrium points and associated phase portraits, thereby revealing the parameter regimes that support distinct classes of nonlinear wave motion. Furthermore, the parameter space analysis establishes an explicit connection between the reduced coefficients and the original physical parameters of the nonlinear Helmholtz equation. Dispersion, carrier modulation, nonparaxiality, Kerr nonlinearity, and self steepening determine the reduced coefficients and corresponding Hamiltonian topology, whereas
	self frequency shift constrains the compatibility manifold required for the real-envelope reduction. Within the same framework, we derived several exact traveling wave solutions, including bright, dark, and periodic waveforms. These results establish a direct connection between the geometry of the reduced phase space and the analytical form of admissible wave solutions, and they demonstrate that self steepening modifies the effective nonlinear coefficient, whereas self frequency shift restricts the admissible real envelope parameter manifold. These mechanisms influence the existence and organization of the resulting wave
	families.
	
	\par To further investigate the nonlinear dynamics of the reduced system, we have examined its response under periodic forcing. The forced system exhibits a transition from regular quasiperiodic oscillations to increasingly complex chaotic dynamics as the forcing amplitude increases. This transition is quantitatively confirmed through the combined use of bifurcation diagrams, largest Lyapunov exponents, and stroboscopic Poincar\'e sections, providing complementary evidence for the forcing-induced transition and a more rigorous characterization of the underlying nonlinear dynamics.
	
	\par More broadly, the present results contribute to the theoretical understanding of nonlinear wave propagation in optical media where paraxial approximations become inadequate. The framework developed here has potential applications for future studies of nonparaxial solitary waves, ultrafast pulse transport, and higher-order nonlinear structures in generalized dispersive systems.
	The analytical traveling wave branches are further validated through direct residual evaluation, numerical propagation of the full nonlinear Helmholtz equation, and robustness tests under $5\%$ and $10\%$ complex Gaussian perturbations. These results demonstrate that the obtained localized and periodic traveling wave solutions remain numerically robust over the tested propagation interval under 5\% and 10\% discrete complex Gaussian perturbations. This supports their computational persistence within the selected parameter and numerical settings of the proposed nonlinear Helmholtz model. Future work may extend the present framework to nonlinear Helmholtz models incorporating higher-order dispersion, Raman scattering, gain-loss mechanisms, saturable or competing nonlinearities, coupled nonlinear Helmholtz equations, and multidimensional or fractional nonparaxial systems. Such extensions would further broaden the applicability of the proposed dynamical systems framework and provide deeper insight into coherent structures and nonlinear wave propagation in realistic optical media.
	
	\section*{AUTHOR DECLARATIONS}
	\subsection*{Conflict of Interest}
	The authors have no conflicts to disclose.
	\subsection*{Author Contributions}
	N.S. and N.K.D. contributed equally to this work.
	N.S., N.K.D., and A.R. conceptualized the research and defined the
	problem; N.S., N.K.D., and A.R. carried out the implementation, A.R. performed numerical analysis, and evaluated
	the results. N.S. and N.K.D, drafted the initial manuscript. A.R. and A.D.
	contributed through revision, editing, feedback, and supervision.
	A.R., N.S. collaboratively finalized the manuscript.

	\section*{DATA AVAILABILITY}
	The data that support the findings of this study are available from the corresponding author upon reasonable request.
	\section*{CODE AVAILABILITY}
	The code is available at the GitHub repository:
	https://github.com/arnob-r/Phase-Space-Reorganization.git.
	\appendix
	\section{CALCULATION OF THE LARGEST LYAPUNOV EXPONENT}\label{appendix1}
	\setcounter{equation}{0}
	\renewcommand{\theequation}{A\arabic{equation}}
	Let
	\[
	\mathbf{x}(\Xi)=
	\begin{pmatrix}
		V(\Xi)\\
		Q(\Xi)\\
		\theta(\Xi)
	\end{pmatrix}
	\]
	denote a reference trajectory, and let
	\[
	\delta\mathbf{x}(\Xi)=
	\begin{pmatrix}
		\delta V(\Xi)\\
		\delta Q(\Xi)\\
		\delta\theta(\Xi)
	\end{pmatrix}
	\]
	denote an infinitesimal perturbation. Linearization of
	Eq.~\eqref{eq:forced_autonomous} about the reference
	trajectory gives
	\begin{equation}
		\frac{d}{d\Xi}\delta\mathbf{x}
		=
		\mathbf{J}\bigl(V(\Xi),\theta(\Xi)\bigr)
		\delta\mathbf{x},
		\label{eq:appendix_variational_general}
	\end{equation}
	where
	\begin{equation}
		\mathbf{J}(V,\theta)=
		\begin{pmatrix}
			0 & 1 & 0\\
			\alpha+3\beta V^{2} & 0 & -\lambda\sin\theta\\
			0 & 0 & 0
		\end{pmatrix}.
		\label{eq:appendix_full_jacobian}
	\end{equation}
	The corresponding variational equations are
	\begin{equation}
		\begin{aligned}
			\delta V' &= \delta Q,\\
			\delta Q' &=
			\left(\alpha+3\beta V^{2}\right)\delta V
			-\lambda\sin\theta\,\delta\theta,\\
			\delta\theta' &= 0.
		\end{aligned}
		\label{eq:appendix_variational_equations}
	\end{equation}
	
	To evaluate perturbation growth at a fixed forcing phase, we
	set
	\begin{equation}
		\delta\theta(0)=0.
	\end{equation}
	Since $\delta\theta'=0$, this condition implies
	$\delta\theta(\Xi)=0$ for all $\Xi$. The tangent dynamics
	therefore reduces to the physical $(V,Q)$ subspace,
	\begin{equation}
		\begin{aligned}
			\delta V' &= \delta Q,\\
			\delta Q' &=
			\left(\alpha+3\beta V^{2}\right)\delta V.
		\end{aligned}
		\label{eq:appendix_reduced_variational_equations}
	\end{equation}
	Although the forcing term does not appear explicitly in
	Eq.~\eqref{eq:appendix_reduced_variational_equations}, it
	affects perturbation growth through the forced reference
	trajectory $V(\Xi)$.
	
	The largest Lyapunov exponent is computed using the
	Benettin tangent-space renormalization
	method~\cite{benettin1980lyapunov}. The reference
	trajectory and one infinitesimal tangent vector are integrated
	simultaneously. In the reduced $(V,Q)$ subspace, define
	\begin{equation}
		\delta\mathbf{y}(\Xi)=
		\begin{pmatrix}
			\delta V(\Xi)\\
			\delta Q(\Xi)
		\end{pmatrix},
	\end{equation}
	with the normalized initial condition
	\begin{equation}
		\delta\mathbf{y}_{0}
		=
		\begin{pmatrix}
			1\\
			0
		\end{pmatrix},
		\qquad
		\left\|\delta\mathbf{y}_{0}\right\|_{2}=1.
		\label{eq:appendix_initial_tangent}
	\end{equation}
	
	The state and tangent equations are integrated over
	successive intervals of length
	\begin{equation}
		T_{\mathrm f}=\frac{2\pi}{\Gamma},
	\end{equation}
	corresponding to one forcing period. At the end of the
	$k$th interval, the tangent-vector norm is
	\begin{equation}
		\rho_k=
		\left\|\delta\mathbf{y}_k\right\|_{2}
		=
		\sqrt{\delta V_k^{\,2}+\delta Q_k^{\,2}}.
		\label{eq:appendix_growth}
	\end{equation}
	Because the tangent vector is normalized at the beginning of
	each interval, $\rho_k$ is the perturbation growth factor over
	that interval. The corresponding logarithmic stretching is
	\begin{equation}
		s_k=\ln\rho_k.
		\label{eq:appendix_log_growth}
	\end{equation}
	The tangent vector is then renormalized according to
	\begin{equation}
		\delta\mathbf{y}_k
		\longleftarrow
		\frac{\delta\mathbf{y}_k}{\rho_k}.
		\label{eq:appendix_renormalization}
	\end{equation}
	This procedure prevents numerical overflow or underflow
	without altering the accumulated exponential growth rate.
	
	The trajectory is first evolved for $N_{\mathrm{tr}}$
	forcing periods to remove transient dependence. During this
	stage, the tangent vector is repeatedly renormalized, but its
	growth factors are not included in the Lyapunov average.
	The largest Lyapunov exponent is then estimated over the
	subsequent $N_{\mathrm m}$ forcing periods as
	\begin{equation}
		\Lambda_{\max}^{(N_{\mathrm m})}
		=
		\frac{1}{N_{\mathrm m}T_{\mathrm f}}
		\sum_{k=1}^{N_{\mathrm m}}
		\ln\rho_k.
		\label{eq:appendix_finite_lyapunov}
	\end{equation}
	The asymptotic exponent is defined by
	\begin{equation}
		\Lambda_{\max}
		=
		\lim_{N_{\mathrm m}\rightarrow\infty}
		\Lambda_{\max}^{(N_{\mathrm m})}.
		\label{eq:appendix_largest_lyapunov}
	\end{equation}
	
	Renormalization once per forcing cycle is not mathematically
	essential, but it ensures that successive stretching factors
	are measured at the same forcing phase and makes the
	Lyapunov calculation directly comparable with the
	stroboscopic Poincar{\'e} section. The Benettin method is
	particularly appropriate here because the governing
	equations and their analytical Jacobian are known explicitly,
	allowing perturbation growth to be obtained directly from the
	variational dynamics without phase space reconstruction or
	nearest-neighbour selection.
	
	A converged positive value of $\Lambda_{\max}$ indicates exponential sensitivity to initial conditions and supports a appearance of chaotic attractor. A value approaching zero is
	consistent with quasiperiodic motion. Numerically, for this study, the dynamics is classified as quasiperiodic when the running largest Lyapunov exponent converges within the numerical tolerance \( |\Lambda_{\max}|<10^{-3} \), with negligible variation under longer integration and smaller time steps. Clearly positive converged values, \( \Lambda_{\max}>10^{-3} \), are classified as chaotic attractor. Small finite-time positive or negative values may nevertheless arise from long transients, intermittency, insufficient
	integration time or numerical error. Convergence is assessed by increasing the transient interval
	and measurement interval, reducing the integration step and
	tightening the solver tolerances. The running estimate
	\[
	\Lambda_{\max}^{(n)}
	=
	\frac{1}{nT_{\mathrm f}}
	\sum_{k=1}^{n}\ln\rho_k
	\]
	should approach a stable plateau as $n$ increases. The
	largest Lyapunov exponent is interpreted together with the
	bifurcation diagram and Poincar{\'e} section. A positive
	converged exponent accompanied by an irregular Poincar{\'e} set
	supports chaotic dynamics, whereas an exponent approaching
	zero together with a smooth invariant curve supports
	quasiperiodic motion.
	
	\section{CONSTRUCTION OF THE POINCAR{\'E} SECTION}\label{appendix2}
	\setcounter{equation}{0}
	\renewcommand{\theequation}{B\arabic{equation}}
	The Poincar{\'e} section provides a discrete representation of the long-time dynamics of the periodically forced system in
	Eq.~\eqref{eq:forced_autonomous}. By recording the trajectory
	at a fixed phase of the external forcing, the explicit forcing
	cycle is removed from the visualization, allowing periodic,
	quasiperiodic and chaotic organization to be distinguished
	more clearly than in a densely filled projection onto the
	$(V,Q)$ plane.
	
	The original periodically forced system is non-autonomous
	because the forcing depends explicitly on $\Xi$. Introducing
	the phase variable
	\[
	\theta=\Gamma\Xi+\theta_0,
	\qquad
	\theta'=\Gamma,
	\]
	converts it into the equivalent three dimensional autonomous
	system evolving in the extended phase space
	\[
	(V,Q,\theta)\in\mathbb{R}^{2}\times S^{1},
	\]
	where $\theta$ denotes the periodic forcing phase. A fixed-phase
	surface of section is defined by
	\begin{equation}
		\Sigma_{\theta_{\mathrm s}}
		=
		\left\{
		(V,Q,\theta)\in\mathbb{R}^{2}\times S^{1}
		:
		\theta=\theta_{\mathrm s}
		\pmod{2\pi}
		\right\},
		\label{eq:appendix_surface_section}
	\end{equation}
	where $\theta_{\mathrm s}$ is the prescribed sampling phase.
	Thus, the three dimensional trajectory is intersected with
	the constant-phase plane
	$\theta=\theta_{\mathrm s}\pmod{2\pi}$ once during each
	forcing cycle. The resulting intersection points are
	projected onto the $(V,Q)$ plane to construct the
	stroboscopic Poincar{\'e} section.
	
	The forcing period is $T_{\mathrm f} = \frac{2\pi}{\Gamma}$. The trajectory intersects
	$\Sigma_{\theta_{\mathrm s}}$ whenever
	\begin{equation}
		\theta(\Xi_n)
		=
		\theta_{\mathrm s}
		\pmod{2\pi}.
		\label{eq:appendix_phase_condition}
	\end{equation}
	The corresponding sampling times are
	\begin{equation}
		\Xi_n
		=
		\Xi_{\mathrm s}+nT_{\mathrm f},
		\qquad
		n=0,1,\ldots,N_{\mathrm P}-1,
		\label{eq:appendix_sampling_times}
	\end{equation}
	where $\Xi_{\mathrm s}$ is the first retained time at the
	chosen forcing phase and $N_{\mathrm P}$ is the number of
	post-transient intersections.
	
	The period-$T_{\mathrm f}$ Poincar{\'e} map is therefore
	\begin{equation}
		\mathcal{M}_{T_{\mathrm f}}
		:
		\begin{pmatrix}
			V_n\\
			Q_n
		\end{pmatrix}
		\longmapsto
		\begin{pmatrix}
			V_{n+1}\\
			Q_{n+1}
		\end{pmatrix},
		\label{eq:appendix_poincare_map}
	\end{equation}
	with
	\begin{equation}
		(V_n,Q_n)
		=
		\bigl(V(\Xi_n),Q(\Xi_n)\bigr).
		\label{eq:appendix_sampled_coordinates}
	\end{equation}
	The plotted Poincar{\'e} set is
	\begin{equation}
		\mathcal{P}
		=
		\left\{
		\bigl(V(\Xi_n),Q(\Xi_n)\bigr)
		\right\}_{n=0}^{N_{\mathrm P}-1}.
		\label{eq:appendix_poincare_set}
	\end{equation}
	
	Only post-transient intersections are retained. If
	$\Xi_{\mathrm{tr}}$ denotes the end of the discarded
	transient interval, the first sampling time is selected such
	that
	\begin{equation}
		\Xi_{\mathrm s}\geq\Xi_{\mathrm{tr}},
		\qquad
		\theta(\Xi_{\mathrm s})
		=
		\theta_{\mathrm s}
		\pmod{2\pi}.
		\label{eq:appendix_first_sampling}
	\end{equation}
	Subsequent sampling at intervals of $T_{\mathrm f}$ preserves
	the same forcing phase. The sampling phase, transient length
	and observation interval are kept fixed when different
	parameter values are compared.
	
	When a uniform integration step $\Delta\Xi$ is used and the
	forcing period contains an integer number of time steps,
	\begin{equation}
		N_{\mathrm f}
		=
		\frac{T_{\mathrm f}}{\Delta\Xi}
		\in\mathbb{N},
		\label{eq:appendix_steps_per_period}
	\end{equation}
	the Poincar{\'e} points can be extracted directly from the
	numerical trajectory using
	\begin{equation}
		j_n
		=
		j_{\mathrm s}+nN_{\mathrm f},
		\qquad
		V_n=V_{j_n},
		\qquad
		Q_n=Q_{j_n}.
		\label{eq:appendix_sampling_indices}
	\end{equation}
	For the present calculations, $\Gamma=2\pi$, so that
	$T_{\mathrm f}=1$. With $\Delta\Xi=0.01$, one forcing period
	contains exactly $N_{\mathrm f}=100$ integration steps.
	Direct index-based stroboscopic sampling is therefore
	phase-consistent.
	
	If $\dfrac{T_{\mathrm f}}{\Delta\Xi}$ is not an integer, the number of
	steps per forcing period should not be obtained by repeated
	rounding. Such rounding can accumulate phase error and
	artificially broaden the section. In that case, the state
	should be evaluated at the exact times
	$\Xi_{\mathrm s}+nT_{\mathrm f}$ using interpolation, or the
	intersections should be located directly through the
	phase-crossing condition in
	Eq.~\eqref{eq:appendix_phase_condition}. Accurate numerical
	localization of intersections is a standard requirement in
	the computation of Poincar{\'e} maps
	\cite{henon1982numerical}.
	
	The geometry of $\mathcal{P}$ characterizes the asymptotic
	forced response. A period-one orbit produces a single
	repeated point, whereas a period-$m$ orbit produces $m$
	distinct points when sampled once per forcing cycle. A
	quasiperiodic response generally produces a smooth invariant
	curve or a family of smooth curves. Chaotic dynamics produces
	an irregular non-smooth set, scattered band or stochastic
	layer, depending on the underlying phase space structure.
	
	A conventional phase portrait contains all points
	$(V(\Xi),Q(\Xi))$ and may appear as a thick band for both
	quasiperiodic and chaotic trajectories. The fixed-phase
	stroboscopic section is more discriminating because the
	forcing phase is held constant. Nevertheless, the Poincar{\'e}
	section alone does not provide an unambiguous diagnosis of
	chaos. Apparent scattering may also arise from unresolved
	transients, high-period motion, insufficient temporal
	resolution or phase-sampling error. The section is therefore
	interpreted together with the largest Lyapunov exponent. A
	smooth invariant curve accompanied by
	$\Lambda_{\max}\simeq0$ supports quasiperiodic motion,
	whereas an irregular Poincar{\'e} set together with a converged
	$\Lambda_{\max}>0$ provides stronger evidence of chaotic
	dynamics.
	
	\section*{References}
	\bibliography{mybibfile}
	\bibliographystyle{apsrev4-1}
	
\end{document}